\documentclass[preprint,12pt,authoryear]{elsarticle}
\usepackage[utf8]{inputenc}

\usepackage[margin=1in]{geometry}
\usepackage{amsmath,amssymb,amsfonts}
\usepackage{tabularx}
\usepackage{graphicx}
\usepackage{textcomp, gensymb}
\usepackage{caption}
\usepackage{subcaption}
\usepackage[hyphens]{url}
\usepackage[hidelinks]{hyperref}
\usepackage[T1,T5]{fontenc}
\usepackage{wasysym}

\hypersetup{breaklinks=true}
\begin{document}
\begin{frontmatter}

\journal{Renewable Energy}
\title{Marine spatial planning techniques with a case study on wave-powered offshore aquaculture farms}

\author[inst1]{Gabriel Ewig\corref{cor1}}
\ead{gre27@cornell.edu}

\author[inst2]{Arezoo Hasankhani}
\ead{arezoo.hasankhani@unh.edu}

\author[inst3]{Eugene Won}
\ead{etw36@cornell.edu}

\author[inst1]{Maha Haji}
\ead{maha@cornell.edu}

\affiliation[inst1]{organization={Sibley School of Mechanical and Aerospace Engineering},
            addressline={Cornell University}, 
            city={Ithaca},
            state={NY},
            postcode={14853}, 
            country={USA}}

\affiliation[inst2]{organization={Department of Electrical and Computer Engineering},
            addressline={University of New Hampshire}, 
            city={Durham},
            state={NH},
            postcode={03824}, 
            country={USA}}            

\affiliation[inst3]{organization={Department of Animal Science},
            addressline={Cornell University}, 
            city={Ithaca},
            state={NY},
            postcode={14853}, 
            country={USA}}

\cortext[cor1]{Corresponding author}

\begin{abstract}
As emerging marine technologies lead to the development of new infrastructure across the ocean, they enter an environment that existing ecosystems and industries already rely on. Although necessary to provide sustainable sources of energy and food, careful planning will be important to make informed decisions and avoid conflicts. This paper examines several techniques used for marine spatial planning, an approach for analyzing and planning the use of marine resources. Using open-source software including QGIS and Python, the potential for developing offshore aquaculture farms powered by a reference model wave energy converter from the Sandia National Labs, the RM3, along the Northeast coast of the United States is assessed and several feasible sites are identified. The optimal site, located at 43.7\degree N 68.9\degree W along the coast of Maine, has a total cost for a 5-pen farm of \$56.8M, annual fish yield of 676 tonnes, and a levelized cost of fish of \$9.23 per kilogram. Overall trends indicate that the cost greatly decreases with distance to shore due to the greater availability of wave energy and that conflicts and environmental constraints significantly limit the number of feasible sites in this region.
\end{abstract}


\end{frontmatter}

%%       \citet{<label>} ==> Jones et al. (2015)
%%       \citep{<label>} ==> (Jones et al., 2015)

\section{Introduction}
\label{sec:introduction}

Oceans have long been a source of opportunity, and conflict, for the communities and industries along their shores. This rich environment, first used as a source of food and for recreation, has since been utilized by other offshore activities including mineral extraction, shipping, aquaculture, and most recently, offshore renewable energy (ORE) systems. Even before the expansion of renewable energy, interactions between different industries led to challenging situations between fisheries and other stakeholders with varying degrees of resolution. \citet{arbo_use_2016}, for example, researched disputes between the oil industry and fisheries in Norway and Vietnam. In Norway, the two industries were able to reach a common agreement to collaborate and work together with an integrated management plan in the Barents Sea, while in Vietnam the two industries have continued to face challenging conflicts. Another study by \citet{andrews_oil_2021} examined the offshore oil industry's adverse impacts on the environment, small-scale fisheries, and coastal community livelihood. Conflicts involving fisheries are further complicated by uncontrolled fishing activities and overlap between recreational and commercial fisheries, which threaten fish stocks and the overall health of the ocean \citep{bess_spatial_2007}.

Entering this mix is the emerging industry of offshore aquaculture, which has the potential to provide a more sustainable and protein-rich source of food for the growing population \citep{noaa_fisheries_offshore_2023}. Compared to conventional aquaculture, offshore aquaculture farms raise fish in the open ocean instead of sheltered bays or onshore tanks. Existing literature by researchers including \citet{gentry_mapping_2017}, \citet{weiss_global_2018}, \citet{brugere_can_2006}, and \citet{bishwajit_fisheries_2014} has brought to light challenges and opportunities for aquaculture farms, including the potential for significant expansion in many parts of the world including Southeast Asia, Australia, North America, and South America. Many of these studies, and particularly \citet{longdill_integrated_2008}, look specifically at the impact on the local environment and how different environmental factors and the siting of farms can change the effects of pollution and other negative disturbances. However, few consider the energy source of the farm, which is often diesel generators, and the potential ways to mitigate the environmental impacts of the associated fossil fuel use.

High ORE growth in response to climate change has created new challenges for many fisheries that currently use the same waters. \citet{stelzenmuller_plate_2022}, studied the impacts of ORE development in the North Sea and the socioeconomic effects on fisheries in that region. Although they find that conflicts exist currently and are ongoing, they predict increasing friction between fisheries and ORE activities after 2025. Some work does show promising signs for the collaboration between existing fisheries and the expanding ORE industry, particularly when coastal communities are brought into the planning process early on. The Block Island Wind Farm, for example, was able to work with local communities and fisheries and decided to change certain plans, including the placement of the turbines, in response to community input \citep{firestone_faring_2020, ten_brink_perceptions_2018}.

The opportunity to co-locate offshore activities together is another promising solution for managing the increasingly congested ocean. Two industries with the potential for co-location are ORE systems and offshore aquaculture farms, where an ORE system can power the farm and potentially shield it from incoming waves. Several studies exist discussing the feasibility of co-locating offshore wind farms to make more efficient use of ocean space including \citet{buck_extensive_2004}, \citet{buck_meeting_2008}, \citet{gimpel_gis_2015}, and \citet{wever_lessons_2015}. Co-location with wave energy converters (WECs) to form a wave-powered aquaculture farm (WPAF) is a particularly interesting option that may be a better match due to a WEC's smaller size and power output, which is closer in scale to the needs of an aquaculture farm. In \citet{garavelli_feasibility_2022}, the authors analyzed the feasibility of this co-location in the United States and identified California and Hawaii as promising sites for future deployment. Co-located WPAFs have also been investigated along the Portuguese coastline, where six different types of WECs have been compared in terms of cost, energy coverage, and capacity factor \citep{clemente_wave_2023}. \citet{whiting_effects_2023} and \citet{silva_effect_2018} have also investigated the near and far-field effects of wave energy which has the potential to dampen waves as they approach an aquaculture farm, although this is still an area of active research.

Previous work by the authors of this paper has developed the groundwork for basic conceptual design and optimization of a WPAF \citep{hasankhani_conceptual_2023} and initial attempts to site such a project in the Northeast United States \citep{hasankhani_marine_2023}. However, a comprehensive framework for marine spatial planning (MSP) using QGIS and Python, specifically one that considers conflicts with other offshore activities and the environmental and geographic requirements for a WPAF, has yet to be detailed. This paper provides a detailed and general framework for the preprocessing and analysis needed for MSP, and applies those techniques alongside an updated cost model to the question of offshore aquaculture siting.

Using these techniques, the authors of this paper aim to assess the potential to integrate wave power into the farm setup, analyze the potential conflicts and environmental constraints that such a farm might face, and provide suggestions for the development of such a farm in the Northeast United States. The farm is modeled using Atlantic salmon, the reference model three (RM3) WEC, and submersible net pens like those from Innovasea, as explained in section \ref{sec:problem-formulation}. By beginning with large-scale mapping techniques, promising sites in the region of interest are identified, and details of the overall considerations that must be made when co-locating wave energy converters and aquaculture farms are revealed. Further integration with Python allows GIS information to be used in a more complex modeling process that includes an objective function to identify cost-effective sites while meeting several constraints. These techniques exemplify the ways that GIS information and modeling could be integrated into a wide range of MSP processes to improve the understanding of the oceanic environment, pick out overall trends relevant to a project, and initiate early engagement with stakeholders to avoid or mitigate potential issues.

Each section of the paper aims to present general information about the MSP process in addition to the specific ways that these methods are applied to the case study of a WPAF. Section \ref{sec:problem-formulation} provides a brief overview of the problem formulation for the application of MSP techniques to a WPAF in the Northeast United States. Section \ref{sec:data-scope} continues with considerations for selecting a scope and data sets for a problem, followed by sections \ref{sec:qgis} and \ref{sec:python-integration} which detail the methods used in both Python and QGIS to process the data sets and model different potential WPAF sites. Finally, sections \ref{sec:results} and \ref{sec:discussion} present the results and discuss observed trends. The Python code referenced in this paper is available in a repository on the SEA Lab GitHub at \href{https://github.com/symbiotic-engineering/aquaculture}{https://github.com/symbiotic-engineering/aquaculture} and the relevant data files are hosted on Zenodo at \href{https://zenodo.org/records/10140826}{https://zenodo.org/records/10140826}.

\section{Problem formulation}\label{sec:problem-formulation}

\begin{figure}[t!]
    \includegraphics[width=1\textwidth]
    {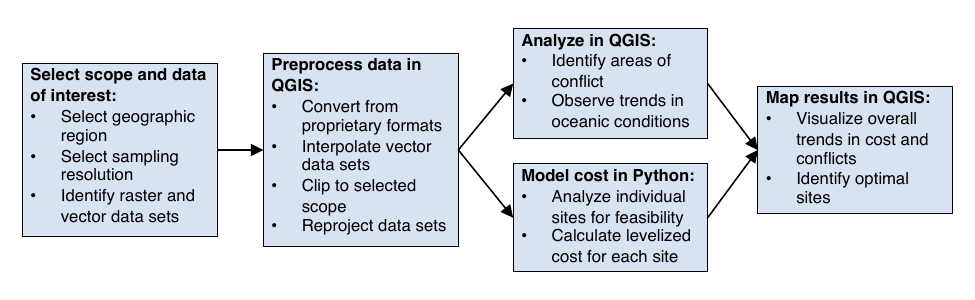}
    \centering
    \caption{Flowchart describing the marine spatial planning process used, including selecting the scope and data, preprocessing data sets, analysis in both QGIS and Python, and visualizing results.}
    \label{fig:msp-flowchart}
\end{figure}

The MSP techniques examined in this paper can be used for a variety of purposes, such as the siting of offshore energy infrastructure and identifying potential conflicts between stakeholders. These techniques are potentially applicable to a wide array of offshore developments, but the specific application of MSP to a WPAF in the Northeast United States is the focus on this study. Figure \ref{fig:msp-flowchart} provides an overview of the process used for this study and other potential applications.

Offshore aquaculture has the potential to provide a sustainable source of seafood for the world's growing population. By taking advantage of more space and higher current speeds further from shore, offshore farms could reduce local environmental impact and expand production. However, the energy demand for necessary equipment like lights and feeders in current designs is supplied by diesel generators. This fossil fuel based source of power further pollutes the environment and requires a regular diesel supply from shore. The use of WECs alongside aquaculture farms has the potential to replace diesel generators and alleviate some of these issues, although many of the local environmental concerns still need to be addressed. In order to investigate this potential co-location, the following points are considered to identify optimal sites in the Northeast United states:

\begin{itemize}
    \item the deployment location should meet the environmental requirements of Atlantic salmon including temperature, salinity, dissolved oxygen, and current speed;
    \item the deployment location should have enough wave power density to power the aquaculture farm, as determined by data on significant wave height and wave energy period;
     \item the WPAF should be installed in a location with a suitable depth for mooring necessary equipment;
     \item the WPAF should avoid conflicts with existing offshore activities such as commercial fisheries and wind lease areas; and
     \item feasible sites should be evaluated for location-dependent costs to minimize the cost of the WPAF.
\end{itemize}

The modeled WPAF includes five cylindrical net pens of 30~m diameter and 15~m height, each with a fish stocking density of $\mathrm{15~\frac{kg}{m^3}}$. The net pen design is similar to those designed by Innovasea, and will be stocked with Atlantic salmon, a common and in-demand species with similar biological needs to steelhead trout, as the target species. The WECs used to power the farm will be the Reference Model 3 (RM3) as designed by Sandia National Laboratories. The RM3 WEC is a two-body heaving point absorber with a top float that moves relative to a vertical column and reaction place when excited by waves. This mechanical energy from ocean waves is converted into electrical energy through a power take-off (PTO) system and a generator within the WEC. The RM3 has a height of 30 m, float diameter of 20m, and a rated capacity of 286 kW \citep{neary_methodology_2014}. Per discussion with aquaculture stakeholders, a spacing of 150 m between the net pens is considered to provide enough space for the mooring and vessel operation. Additionally, the spacing between WECs is considered twice the WEC diameter to avoid any destructive wave interaction and reduction in the generated power \citep{zhong_wave-body_2019}. Two feed barges are also required to enable automatic fish feeding; each barge has six feed lines and a capacity of 200 metric tons \citep{scale_aquaculture_as_seafarm_2019}. Travel to the net pens using ships is scheduled weekly to restock the feed barges and check the aquaculture farm.

\begin{figure}
    \centering
    \begin{subfigure}[b]{0.3\textwidth}
        \centering
        \includegraphics[width=\textwidth]{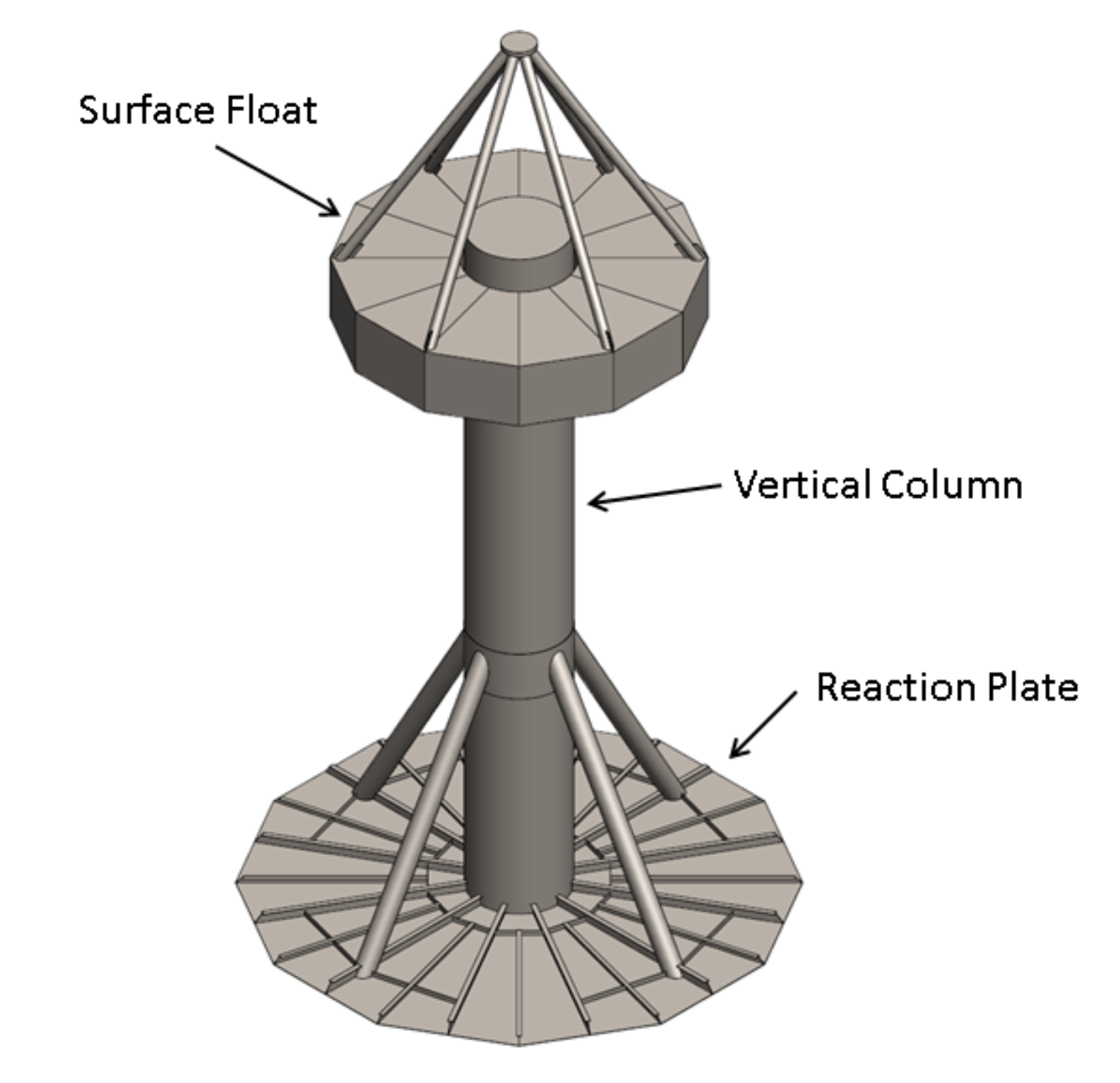}
        \caption{RM3 WEC}
        \label{fig:rm3-wec}
    \end{subfigure}
    \begin{subfigure}[b]{0.45\textwidth}
        \centering
        \includegraphics[width=\textwidth]{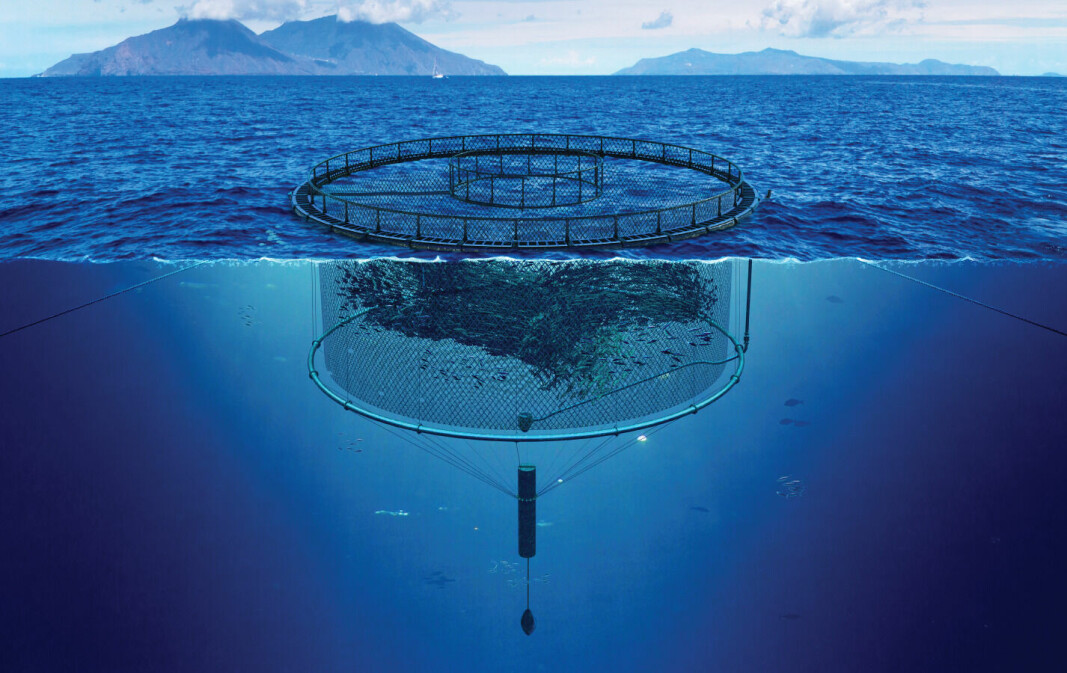}
        \caption{Innovasea submersible net pen}
        \label{fig:net-pen}
    \end{subfigure}
    \caption{Reference model 3 (RM3) wave energy converter \citep{neary_methodology_2014}, and Innovasea submersible net pen considered in this study \citep{innovasea_innovasea_2023}.}
    \label{fig:net-pen-and-wec}
\end{figure}

\section{Scope and data selection}\label{sec:data-scope}
\subsection{Study scope} \label{subsec:study-scope}

\begin{figure}[t!]
    \includegraphics[width=\textwidth]
    {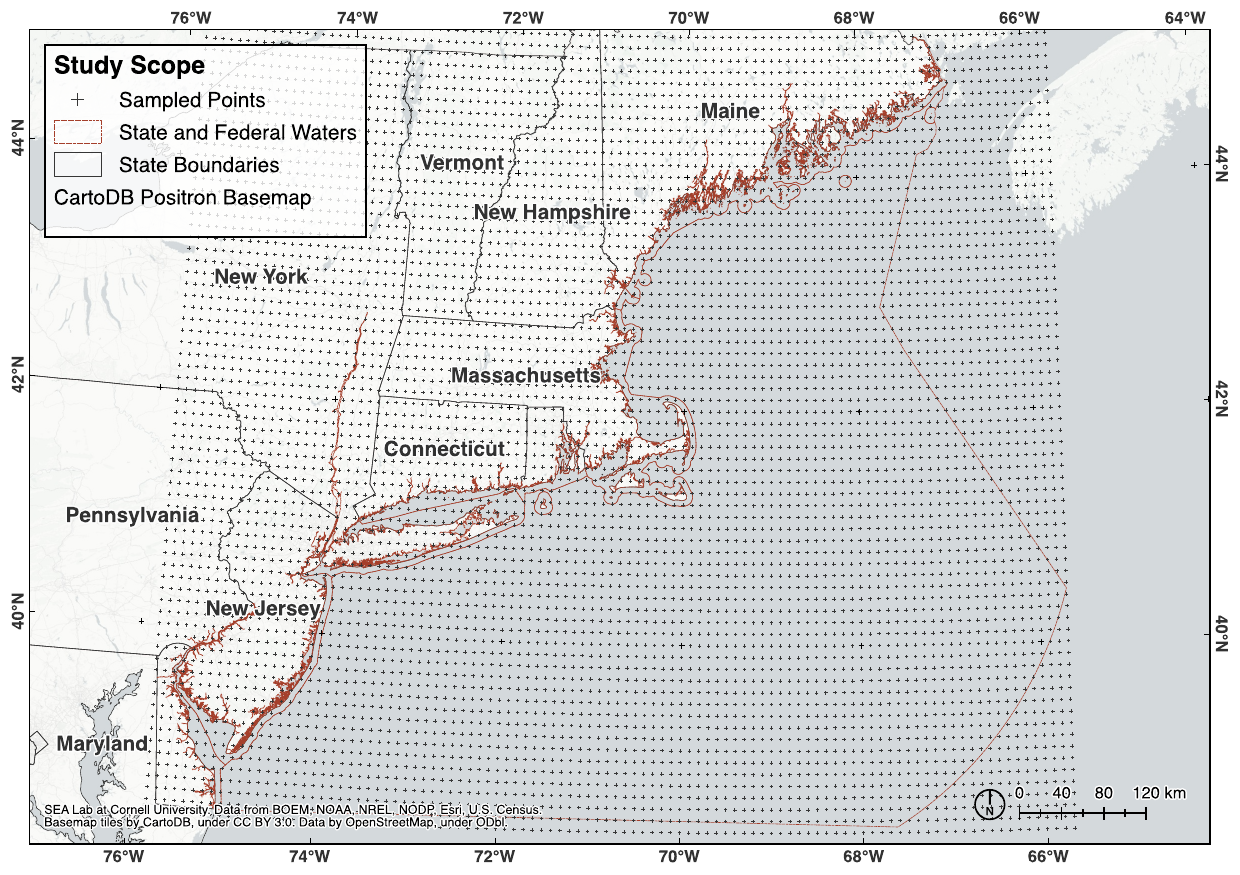}
    \centering
    \caption{Selected region of Northeastern United States including state and federal waters, along with the {0.1\degree} grid of sampled points (points over land are excluded early in analysis).}
    \label{fig:study-scope}
\end{figure}

This study focuses on waters along the Atlantic coast of the Northeastern United States. The region is home to several large fisheries, and the industry holds important historical and cultural significance in the area. Overfishing and environmental changes over the past decades, however, have greatly diminished fish stocks leading to record lows in annual catches of cod and other species \citep{whittle_haul_2022}. The region is also home to several aquaculture operations, although the authors are not aware of any currently located far offshore as suggested in this study. Northeast Sea Grant defines the Northeast as the states of Maine, New Hampshire, Massachusetts, Rhode Island, Connecticut, and New York \citep{northeast_sea_grant_northeast_2022}, which are therefore the main focus of this study. However, to include all regions that are easily accessible from ports in these states, this study uses a slightly larger scope which also includes Delaware and New Jersey. The study includes both state waters, within 5.6km (3 nautical miles) of shore, and federal waters, within 370km (200 nautical miles) of shore, as shown in figure \ref{fig:study-scope}. This region is roughly bounded by 38.40\degree N 75.80\degree W and 45.20\degree N 65.70\degree W.

There are several approaches to evaluating complex models across a large region, depending on the needs of a project. Projects aiming to identify individual optimal sites may benefit from optimization techniques such as gradient descent that search for locations that minimize an objective function such as cost. For this study, however, a brute-force approach is used to allow the entire region to be evaluated so that overall trends can easily be identified. Unlike an optimization approach which strategically searches for a minimum, the brute-force approach evaluates every location at a set interval across the region. Although more time-consuming, the results of this approach provides a more complete understanding of the entire region. The results using this approach are achieved in a reasonable time frame, under an hour, for the current model, however a more complex model may necessitate a different approach to optimization. The interval at which sites are evaluated with this approach determines the ``resolution'' of the results, which is limited by how much processing time is practical and the lowest resolution of the data sets used in the study. Some common properties like bathymetry have very high-resolution data sets available, while others may be lower resolution or limited to smaller areas. For this study, a resolution of {0.1\degree} longitude and latitude (roughly 10km in this region) is selected to balance the need for precision while not being overly demanding of computational time. This is slightly lower than the most limiting data sets of wave conditions, which have a resolution of about {0.067\degree}. Figure \ref{fig:study-scope} shows the points sampled at this resolution alongside the state and federal waters included in the study scope. While this resolution provides general trends across the entire region, a higher resolution pass would be necessary for later stages of planning, especially once a more specific region of interest can be identified. Table \ref{table:data-sources-raster} includes the resolution of individual data files, which indicate the highest resolution that could be achieved before other data sources become necessary.

\subsection{Sources of data}
\label{sec:sources-of-data}

GIS data types fall into two general types: raster and vector data sets. Raster data sets provide pixels of continuous information like physical conditions or satellite imagery, while vector data sets describe points, lines, or areas. In this study, raster data sets are used for oceanic conditions and bathymetry, while vector data sets describe regions of potential conflict and administrative boundaries. Current speed, dissolved oxygen, temperature, and salinity are used to determine the carrying capacity. Significant wave height and wave energy period are used to determine the wave power potential, which along with distance to port is used in the cost model. Bathymetry data, regions with other active industries, and restricted areas are used to constrain the feasible area. Together, these data sets are used to evaluate the location-dependent costs for each site and identify potential conflicts, as discussed in sections \ref{sec:cost-function} and \ref{sec:constraints}.

The data in this case study is sourced from several organizations that have publicly available GIS data including the Bureau of Ocean Energy Management (BOEM), the National Oceanic and Atmospheric Administration (NOAA), the National Renewable Energy Lab (NREL), the Northeast Ocean Data Portal (NODP), and the United States Census. Tables \ref{table:data-sources-raster} and \ref{table:data-sources-vector} contain an overview of the information used in the case study including the data type and file source. Section \ref{sec:preprocessing-data} explains the steps taken to standardize these different data sets which includes file conversion, interpolation, clipping, and re-projection. Preprocessed versions of each of these data sets are hosted on Zenodo at \href{https://zenodo.org/records/10140826}{https://zenodo.org/records/10140826}.

\begin{table}[t!]
\footnotesize
\caption{Raster data sets that are included in the study, including those relating to environmental conditions, wave energy potential, bathymetry, and distance to port. Source and approximate resolution are also noted.}
\begin{center}
\begin{tabularx}{\textwidth}{| >{\centering\arraybackslash}X | >{\centering\arraybackslash}c | >{\centering\arraybackslash}X | >{\centering\arraybackslash}X |}
 \hline
 \textbf{Name} &  \textbf{Unit} &  \textbf{Notes} &  \textbf{Source}\\ %[0.5ex] 
 \hline Current speed & m/s & Finite volume community ocean model (FVCOM) annual climatology, {0.002\degree} resolution downsampled to {0.004\degree} & \cite{northeast_ocean_data_portal_fvcom_2016} \\ 
 \hline Dissolved oxygen & mg/L & Annual climatology, interpolated from vector data with {1\degree} resolution to a raster with {0.015\degree} resolution & \cite{national_centers_for_environmental_information_world_2019}\\ 
 \hline Salinity & PSU & Annual climatology, interpolated from vector data with an irregular but roughly {0.1\degree} resolution to a raster with {0.01\degree} resolution& \cite{national_centers_for_environmental_information_world_2019}\\ 
 \hline Temperature & \degree C & FVCOM annual climatology, {0.002\degree} resolution & \cite{northeast_ocean_data_portal_fvcom_2016}\\ 
 \hline Wave height & m & 51-month Wavewatch III hindcast, {0.067\degree} resolution & \cite{national_renewable_energy_laboratory_marine_2011}\\ 
 \hline Wave period & s & 51-month Wavewatch III hindcast, {0.067\degree} resolution & \cite{national_renewable_energy_laboratory_marine_2011}\\ 
 \hline Bathymetry & m & {0.0007\degree} resolution, downsampled to {0.003\degree} & \cite{national_geophysical_data_center_bathymetry_1990}\\ 
 \hline Distance to port & m & {0.005\degree}vector generated from vector of principal ports & \cite{office_for_coastal_management_principal_2019}\\ 
 \hline
\end{tabularx}
\end{center}
\vspace{-5mm}
\label{table:data-sources-raster}
\end{table}

\begin{table}[t!]
\footnotesize
\caption{Vector data sets that are included in the study, including those relating to conflicting industries, restricted areas, and administrative boundaries.}
\begin{center}
\begin{tabularx}{\textwidth}{| >{\centering\arraybackslash}X | >{\centering\arraybackslash}X | >{\centering\arraybackslash}X |}
 \hline
  \textbf{Name} &  \textbf{Notes} &  \textbf{Source}\\ %[0.5ex] 
 \hline Fishing traffic & Generated using raster from automatic identification system (AIS) transponders & \cite{northeast_ocean_data_portal_fishing_2022}\\
 \hline Marine protected areas & & \cite{national_marine_protected_areas_center_mpa_2020}\\
 \hline Military zones & Includes danger zones, restricted areas, submarine transit lanes, and testing areas & \cite{northeast_ocean_data_portal_national_2016, office_for_coastal_management_danger_2022} \\
 \hline Offshore wind & Includes wind leases, planning areas, and existing farms & \cite{northeast_ocean_data_portal_wind_2015,bureau_of_ocean_energy_management_renewable_2023,northeast_ocean_data_portal_rhode_2010}\\
 \hline Shipping lanes & & \cite{office_of_coast_survey_shipping_2015}\\
 \hline State, federal waters & & \cite{office_for_coastal_management_federal_2018}\\ 
 \hline USA major cities & & \cite{esri_usa_2023}\\ 
 \hline
\end{tabularx}
\end{center}
\vspace{-5mm}
\label{table:data-sources-vector}
\end{table}

It is important to select data that is a good fit for the study, especially when multiple sources are available. With significant wave height and wave period, for example, an older raster data set of annual averages from NREL is chosen because it provides continuous information across the region and is available in a format that is readily compatible with GIS. NOAA's National Data Buoy Center, another possible option, could have provided the same information but would have limited the analysis to the few locations where buoys already exist and would have required more processing to integrate into the model. In other cases, available data sets may be limited by resolution or region. An older bathymetry data set, for example, is selected because it provides a high resolution across the entire study scope while more recent surveys are limited to a smaller region.

\subsection{Software selection}

Software is selected for this work with the intention of being accessible to a wide range of applications and users who might benefit from MSP. As a result, only open-source software is used in the final process. GIS software is used for basic processing and visualization of results including the layout of published maps. QGIS is used as the main GIS software because of its widespread availability and complete set of processing tools. QGIS unfortunately lacks the ability to process certain proprietary data types from ArcGIS, the current industry standard, but is otherwise a capable and free option. Python, along with the GeoPandas package, is used to handle more complex modeling such as the carrying capacity and cost functions which would be difficult in GIS software alone. For these cases, a Python package is created to interface between the main model and the individual GIS data sets, as demonstrated in section \ref{sec:python-integration}.

\section{QGIS processing and analysis}\label{sec:qgis}
\subsection{Preprocessing data}\label{sec:preprocessing-data}

Several preprocessing steps are necessary to standardize downloaded GIS data files and prepare them for further mapping and analysis. These include occasional file conversion to a QGIS-compatible format, vector interpolation for environmental conditions, clipping data sets to the study scope, and reprojection to either WGS 84 or WGS 84 / UTM Zone 19. More thorough instructions to complete these steps in QGIS are provided in \ref{appendix:qgis-instructions}.

\subsection{Raster analysis} \label{sec:raster-analysis}

\begin{figure}[t!]
    \includegraphics[width=\textwidth]{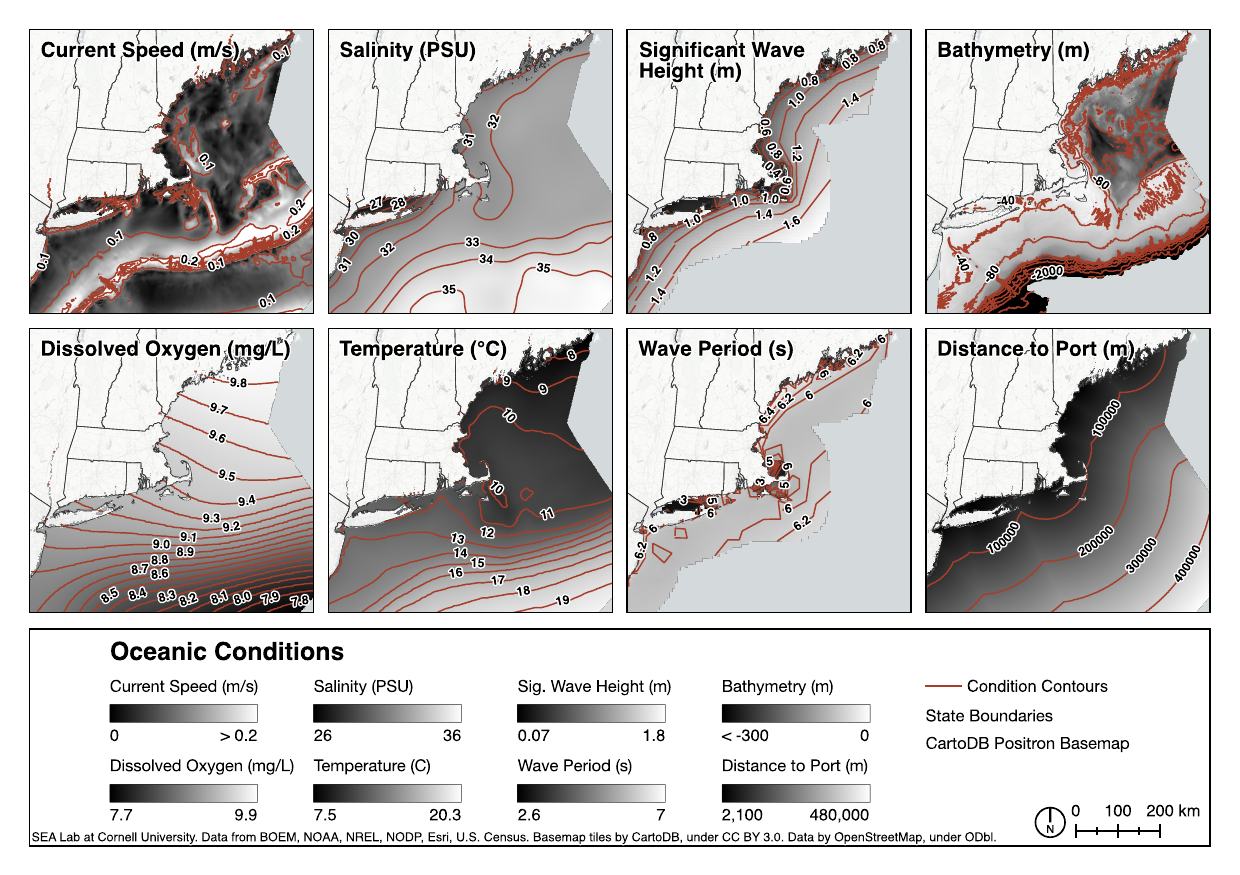}
    \centering
    \caption{Raster data sets of conditions considered in the analysis with generated contours overlaid. Environmental conditions and bathymetry are used to identify feasible sites, while the wave climate and distance to port are used in the cost model.}
    \label{fig:oceanic-conditions}
\end{figure}

Although most analysis for this case study is done within the Python model shown in section \ref{sec:python-integration}, there are several helpful functions in QGIS that can aid in understanding and visualizing the data. For this study, two are used: generating contours to display the range of oceanic conditions, and highlighting regions that meet certain conditions or constraints with Raster Calculator. Figure \ref{fig:oceanic-conditions} shows all raster data sets used in the case study, including environmental conditions used to identify favorable conditions for a certain species, average wave period and height to estimate WEC power output, and bathymetry and distance to port which further constrain feasible sites and affect the cost. The environmental conditions are also used to evaluate the carrying capacity of a location, which indicates how large a population can be sustained in a given environment. Generating contours and using the Raster Calculator in QGIS are both discussed in \ref{appendix:qgis-instructions}.

\subsection{Vector analysis}\label{sec:vector-analysis}

\begin{figure}[t!]
    \includegraphics[width=\textwidth]{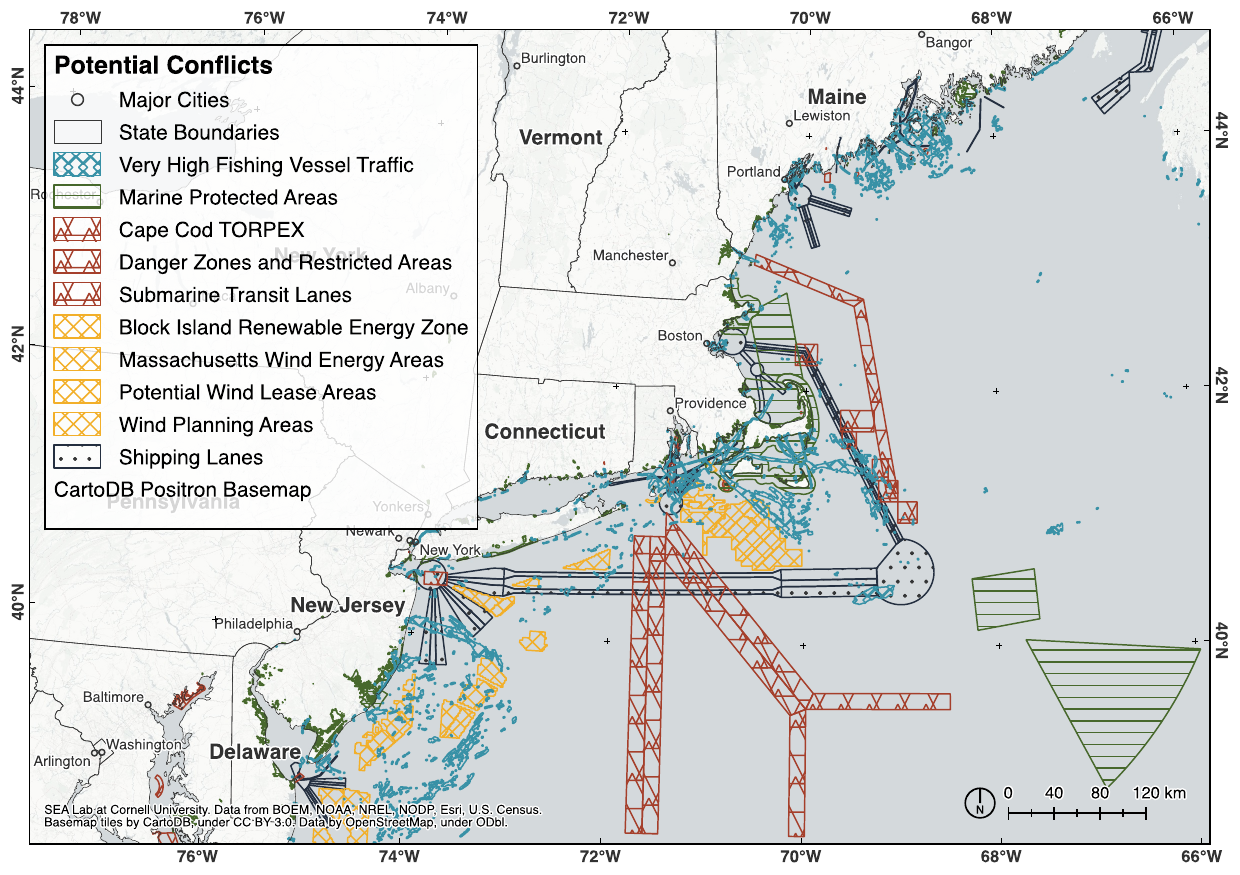}
    \centering
    \caption{Data sets for all regions with conflicting industries or restricted areas that are excluded from the analysis, including shipping lanes, marine protected areas, and more.}
    \label{fig:potential-conflicts}
\end{figure}

In this study vector data is used to represent conflicts, which QGIS can aid in identifying. It is often helpful to convert between raster and vector formats depending on the analysis needed. Areas with high fishing vessel traffic, for example, are excluded in this analysis and so a vector of regions that have high fishing traffic is first created using the raster input data. Vector data sets are the core of the conflict analysis used to identify overlapping industries, and each of the included data sets are pictured in figure \ref{fig:potential-conflicts}. 

\begin{figure}[t!]
    \includegraphics[width=\textwidth]{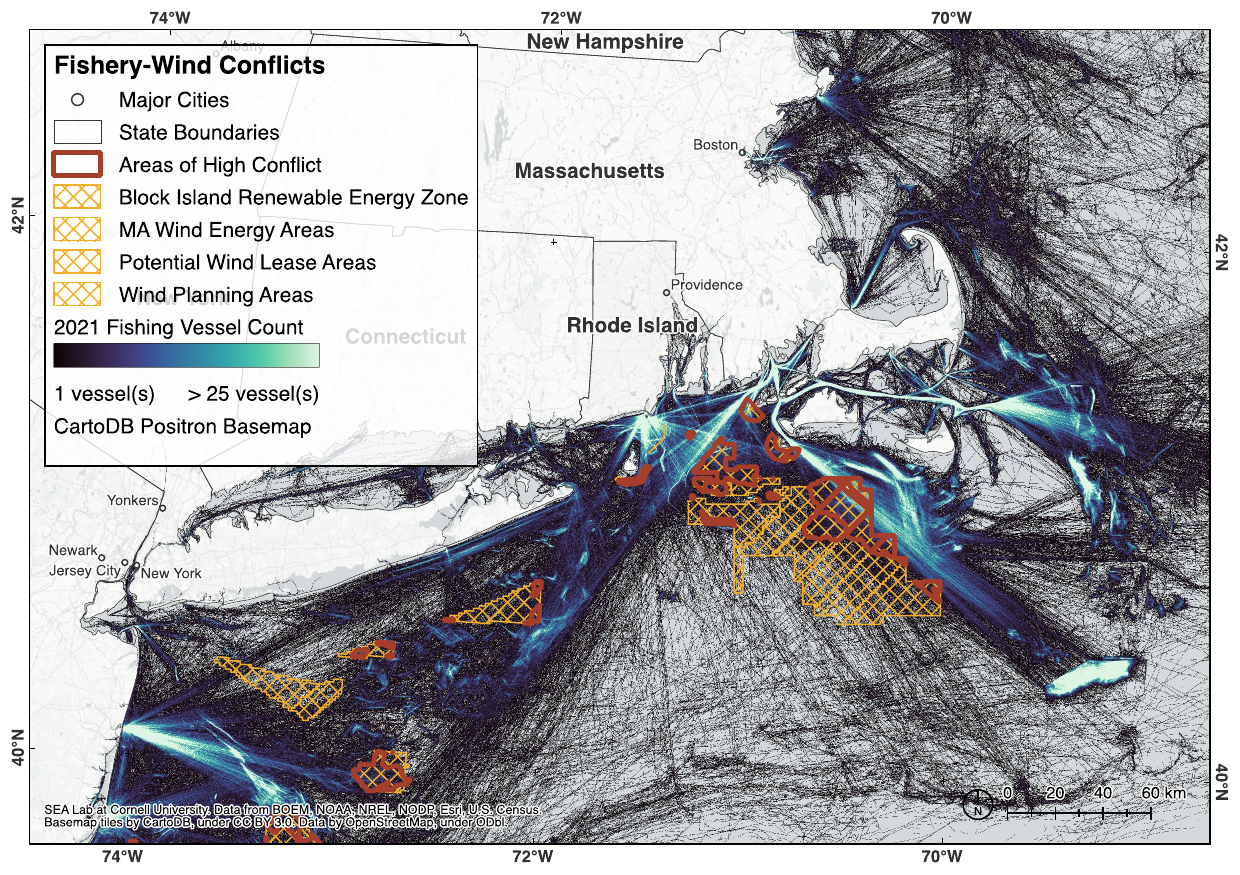}
    \centering
    \caption{Fishing vessel traffic and offshore wind lease areas near Rhode Island with regions of overlap, and therefore a potential for conflict, highlighted.}
    \label{fig:fishing-conflicts}
\end{figure}

Like raster files, vector data sets are later loaded into Python, but preliminary maps of conflicts can also be generated in QGIS to identify and highlight potential issues early on. Figure \ref{fig:fishing-conflicts} demonstrates this technique using the data sets for wind lease areas and regions of high fishing traffic, and highlights the regions where these industries overlap and therefore could conflict. Note that although the regions highlighted in figure \ref{fig:fishing-conflicts} include areas that had above-average fishing traffic in 2021, only areas with very high fishing traffic (over one standard deviation above the average) are considered infeasible farm locations in the Python model. This conclusion was determined after attending meetings hosted by the Bureau of Ocean Energy Management with stakeholders in the fishing and offshore wind industries. The fishing community is very protective of their fishing grounds and information about where they are is limited. Therefore, this assumption reflects a reasonable understanding of potential conflicts with fisheries and other industries.

\section{Python integration and modeling} \label{sec:python-integration}

\subsection{GIS-Python interface}\label{gis-python-interface}
\begin{figure}[t!]
    \includegraphics[width=\textwidth]{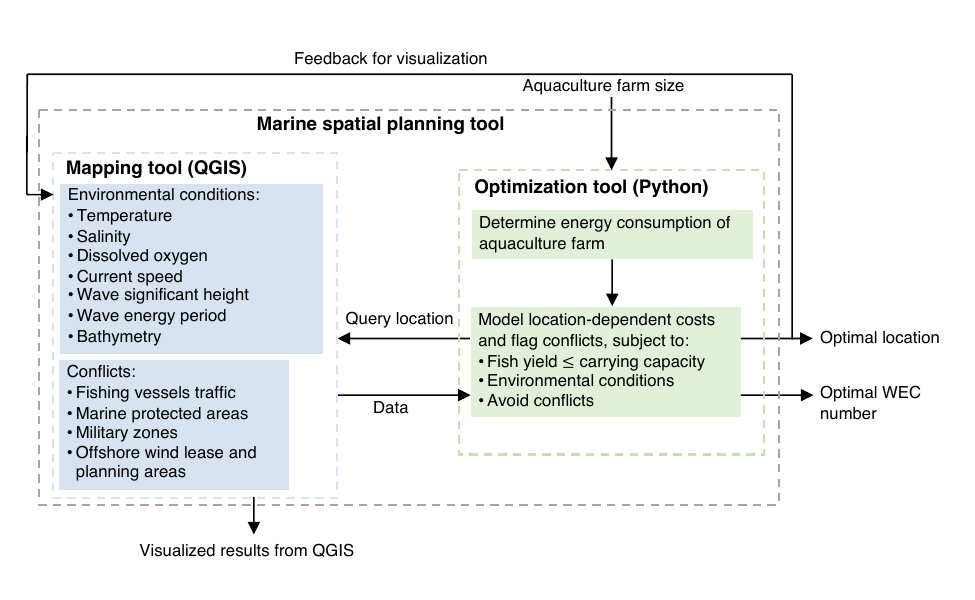}
    \centering
    \caption{The analysis process between QGIS and Python for the marine spatial planning process.}
    \label{fig:software-flowchart}
\end{figure}

Although QGIS is a powerful tool for the analysis and visualization of geospatial data, some functions like modeling and optimization are difficult in QGIS alone. For this reason, Python is used as an intermediary step for more complex analysis. Figure \ref{fig:software-flowchart} shows the overall process for using both approaches where the Python optimization tool uses a ``handler'' object to query geospatial data processed in QGIS to evaluate the location-dependent costs of the farm. This handler and examples of its usage are available on the SEA Lab GitHub at \href{https://github.com/symbiotic-engineering/aquaculture}{https://github.com/symbiotic-engineering/aquaculture}.

After completing the basic preprocessing described in \ref{sec:preprocessing-data}, GeoPandas, Shapely, and Rasterio packages are used to process geospatial data in Python. First, a custom "handler" object is created in Python to handle the integration of the geospatial data sets with the Python model, which expects simple numerical values for conditions and flags for when there are potential conflicts. The handler first constructs a GeoPandas GeoDataFrame, which stores data and the results from the Python model in a format that can later be read again by QGIS. Conflict vectors are also loaded as separate GeoDataFrames, while condition rasters are loaded using Rasterio. 

The handler allows the model to query individual points by latitude and longitude and passes along the value at that point for each loaded condition raster. It then flags the point if overlaps with any of the vectors that have been loaded. One flag is used to identify conflicts, another is used to indicate whether the point queried is out of the study scope, and a third is used if all raster condition data sets exist at the given point. The handler then receives the result for a given point from the cost model and stores it in a GeoDataFrame. Although individual queries are helpful for an optimization approach, the handler can also query information for many points at once to fully populate a grid of potential sites. Finally, the GeoDataFrame with the model result is exported again as a GeoJSON file that can be loaded into QGIS to create final maps of the results.

\subsection{Cost model}\label{sec:cost-function}

The cost model evaluates the costs associated with development of a WPAF at a given site, and is modeled entirely in Python. A brief explanation of the cost model as relevant to the WPAF case study is included here, but details about this function and the optimization process can be found in a previous paper by this author, \citet{hasankhani_marine_2023}. In this study, the objective is to minimize the location-dependent cost of the WPAF, and the design variable is the WPAF deployment location which determines the environmental and geographic conditions used as inputs in the model.

The two WPAF cost terms that are the most dependent on the deployment location include the cost of the WECs and the cost of vessel travel associated with weekly trips to check the farm and restock the automatic feeders. The cost of WECs is determined by the availability of wave energy at a location and, from that, the number of RM3 WECs necessary to power a farm of a given size. The energy yield of the WEC is calculated considering the wave power density of the WPAF site and the capture width, which is the ratio of the power absorbed by the WEC to the wave power density of the site. This capture width is defined as the product of the physical dimension of the WEC and the hydrodynamic efficiency of the WEC, known as the capture width ratio. The capture width for an array of WECs is considered to be the sum of each individual WEC's capture width for this analysis. The wave power is found through the annual average of the significant wave height and wave period, as discussed in section \ref{sec:sources-of-data}. The model does not take into account the layout, direction, or near and far-field effects of the WECs and pens. Vessel travel cost to the WPAF is the second operating expense which increases for deployment locations further from the coast due to higher travel costs. This vessel travel cost is defined by the necessary travel time multiplied by the cost of labor and fuel. Though the total cost of the WPAF system depends on additional factors such as net pen design cost, feed barge cost, fish feed cost, and fingerling cost, these remain constant for a designated size of aquaculture farm (e.g. five cylindrical net pens measuring 15~m in height and 30~m in diameter). These constant costs are therefore not modeled in this case study, but are the subject of future work. The location-dependent costs are modeled over the total lifetime of the project, including capital expenditures (CapEx) and operating expenses (OpEx). The objective function and total cost formulation is presented in equations \ref{eq:objective-function} and \ref{eq:total-cost} where $r=0.07$ is the discount rate \citep{femp_2021_2021} and $t$ reaches the expected lifetime of 15 years.

\begin{equation}\label{eq:objective-function}
    \min_{\mathrm{location}} \mathrm{total~cost}_{\mathrm{WEC}} + \mathrm{total~cost}_{\mathrm{vessel~travel}}
\end{equation}
\begin{equation}\label{eq:total-cost}
    \mathrm{total~cost}= \mathrm{CapEx} + \sum_{t=1}^{15} \frac{\mathrm{OpEx}}{(1+r)^{t}}
\end{equation}

The CapEx and OpEx terms for WECs are defined by multiplying the number of WECs required to power the farm of a set size by the reference CapEx and OpEx of each RM3 WEC. These reference CapEx and OpEx for the RM3 are determined considering the economies of scale as detailed in \citep{neary_methodology_2014}.  Values for costs that are not location-dependent are referenced in the results section in table \ref{table:optimal-site-costs}. Constant capital expenditures include the net pen and feed barge, while constant operating expenses include fish feed and the fingerlings used to stock the farm at the beginning of each season. For this study, a harvest period of one year is considered, where the fingerlings are put in the net pens at a weight of 200g. Note that the cleaning costs of the net pens and feed barge are considered zero since they are negligible compared to the capital cost \citep{innovasea_innovasea_2023}.

A metric like levelized cost of energy is also used to calculate the cost per fish yield of a farm at a given site. This levelized cost of fish is calculated by dividing the total farm cost by the discounted fish yield over the lifetime of the farm, as shown in equation \ref{eq:cost-per-fish-yield}  where $r$ and $t$ are the same as in equation \ref{eq:total-cost}.

\begin{equation}\label{eq:cost-per-fish-yield}
    \mathrm{levelized~cost~of~fish} = \frac{\mathrm{total~cost}}{\sum_{t=1}^{15} \frac{\mathrm{fish~yield} }{(1+r)^{t}}}
\end{equation}

It should also be mentioned that the cost analysis in this study is made per an assumption that the energy demand of the aquaculture farm is modeled as a constant annual value using a reference value for power per fish yield reported by \cite{freeman_offshore_2022}. Hence, the energy demand is overestimated and considered constant per year, and hourly fluctuation of the power consumption of the aquaculture farm, such as an increase in energy demand while fish feeding, is not considered in this study. Furthermore, the uncertainty of the wave power, (i.e. the hourly data of wave significant height and wave energy period) is not considered in this study, and all the results are presented with the annual average wave data. Therefore, the cost of the WPAF, and specifically the cost of WEC, could change in future studies using high-resolution hourly data for aquaculture farm energy demand and wave power.

\subsection{Constraints}\label{sec:constraints}

The analysis assess for a variety of constraints to exclude sites that do not meet the feasibility requirements for WPAF development. These constraints include ensuring that a given site does not intersect with other offshore industries or restricted areas, that the necessary carrying capacity for Atlantic salmon is met, and that a suitable depth and environmental conditions within an acceptable range are present. These constraints are formulated in equations \ref{eq:conflict-constraints}, \ref{eq:fish-constraints}, and \ref{eq:env-constraints} where $CC_{\mathrm{O_2}}$ refers to the carrying capacity considering oxygen availability.

\begin{equation}\label{eq:conflict-constraints}
\mathrm{exclude~conflicts~with}
\begin{bmatrix}
\mathrm{fishing~vessels}\\
\mathrm{shipping~lanes}\\
\mathrm{military~zones}\\
\mathrm{marine~protected~areas}\\
\mathrm{offshore~wind~farms}\\
\end{bmatrix}
\end{equation}
\begin{equation}\label{eq:fish-constraints}
    \mathrm{fish~yield} \leq \mathrm{CC}_\mathrm{O_{2}}
\end{equation}
\begin{equation}\label{eq:env-constraints}
%\resizebox{0.92\textwidth}{!}{$    
\begin{bmatrix}\mathrm{temperature}_{\mathrm{min}}\\\mathrm{salinity}_{\mathrm{min}}\\\mathrm{dissolved~oxygen}_{\mathrm{min}}\\\mathrm{current~speed}_{\mathrm{min}}\\ \mathrm{depth}_{\min}\end{bmatrix} 
    \leq  
    \begin{bmatrix}\mathrm{temperature}\\ \mathrm{salinity} \\ \mathrm{dissolved~oxygen} \\ \mathrm{current~speed} \\ \mathrm{depth}\end{bmatrix} 
    \leq \begin{bmatrix}\mathrm{temperature}_{\mathrm{max}}\\\mathrm{salinity}_{\mathrm{max}}\\\mathrm{dissolved~oxygen}_{\mathrm{max}}\\\mathrm{current~speed}_{\mathrm{max}}\\ \mathrm{depth}_{\max} \end{bmatrix}
   % $}
\end{equation}

Intersection with existing industries and protected areas is evaluated directly by the GIS handler using the vector data sets of these regions. This information is then passed along with the environmental conditions from the raster GIS files at a selected site to the python model which evaluates the carrying capacity and environmental constraints alongside the objective function of location-dependent costs.

The carrying capacity is a measure of the amount of fish biomass that can be placed in the aquaculture farm while maintaining a healthy environment and is in this study determined using the current speed and dissolved oxygen \citep{stigebrandt_regulating_2004}. The carrying capacity of a row farm is formulated in equation \ref{eq:carrying-capacity} where $\mathrm{O}_{\mathrm{2in}}$ is the average dissolved oxygen at the deployment site, $\mathrm{O}_{2\min}$ is the minimum dissolved oxygen required for the fish, $D_{\mathrm{pen}}$ is the net pen diameter, $H_{\mathrm{pen}}$ shows the net pen height, $U_{\min}$ is the minimum current speed, $Pe$ is the permeability of the farm, and $\mathrm{DO}_2$ is the mean oxygen consumption rate per kilogram of fish production per day.

\begin{equation}\label{eq:carrying-capacity}
%\resizebox{0.43\textwidth}{!}{$
\mathrm{CC}_{\mathrm{O_{2}}}=\frac{86400(\mathrm{O}_{\mathrm{2in}} - \mathrm{O}_{2\min}) n_{\mathrm{pens}} D_{\mathrm{pen}} H_{\mathrm{pen}} U_{\min} Pe}{\mathrm{DO_2}}
%$}
\end{equation}

Finally, the feasibility of the environmental conditions and bathymetry are evaluated for a given site. Table \ref{table:data-bounds} lists the range of acceptable values for each of these constraints. The bathymetry constrain is set by the feasibility of mooring submersible net pens to the ocean floor, but the connections and power distribution between the WECs, net pens, and other equipment are not considered. The mooring requirements of the RM3 WEC of 40-100m \citep{neary_methodology_2014} are considered but the requirements for submersible net pens of 40-75m are more limiting.

\begin{table}[!th]
    \footnotesize
    \caption{Constraints for a sample offshore Atlantic salmon WPAF}
      \vspace{-5mm}
    \begin{center}
    \begin{tabular}{|c| c | c | c | c |} 
     \hline
     \textbf{Parameters} & \textbf{Description} & \textbf{Unit} & \textbf{Value} & \textbf{Source}\\ 
     \hline
        $T_{\mathrm{min}}$ & Minimum temperature & $^\circ\mathrm{C}$ & 2 & \cite{pecherska_us_2019} \\ 
     \hline 
        $T_{\mathrm{max}}$ & Maximum temperature & $^\circ\mathrm{C}$ & 20 & \cite{pecherska_us_2019} \\ 
     \hline 
        $Sa_{\mathrm{min}}$ & Minimum salinity & PSU & 30 & \cite{pecherska_us_2019} \\ 
     \hline 
        $Sa_{\mathrm{max}}$ & Maximum salinity & PSU & 35 & \cite{pecherska_us_2019} \\ 
     \hline
        $O_{2\min}$ & Minimum dissolved oxygen & mg/L & 4.41 & \cite{pecherska_us_2019} \\ 
     \hline 
        $O_{2\max}$ & Maximum dissolved oxygen & mg/L & - & \\ 
     \hline
        $U_{\min}$ & Minimum current speed & m/s & 0.01 & \cite{pecherska_us_2019} \\ 
     \hline
        $U_{\max}$ & Maximum current speed & m/s & 2 & \cite{pecherska_us_2019} \\ 
     \hline
        $de_{\min}$ & Minimum depth & m & 40 & Stakeholders \\ 
     \hline
        $de_{\max}$ & Maximum depth & m & 75 & Stakeholders \\ 
     \hline
    \end{tabular}
    \end{center}
      \vspace{-5mm}
    \label{table:data-bounds}
\end{table}

\section{Results}\label{sec:results}
\begin{figure}[t!]
    \includegraphics[width=\textwidth]{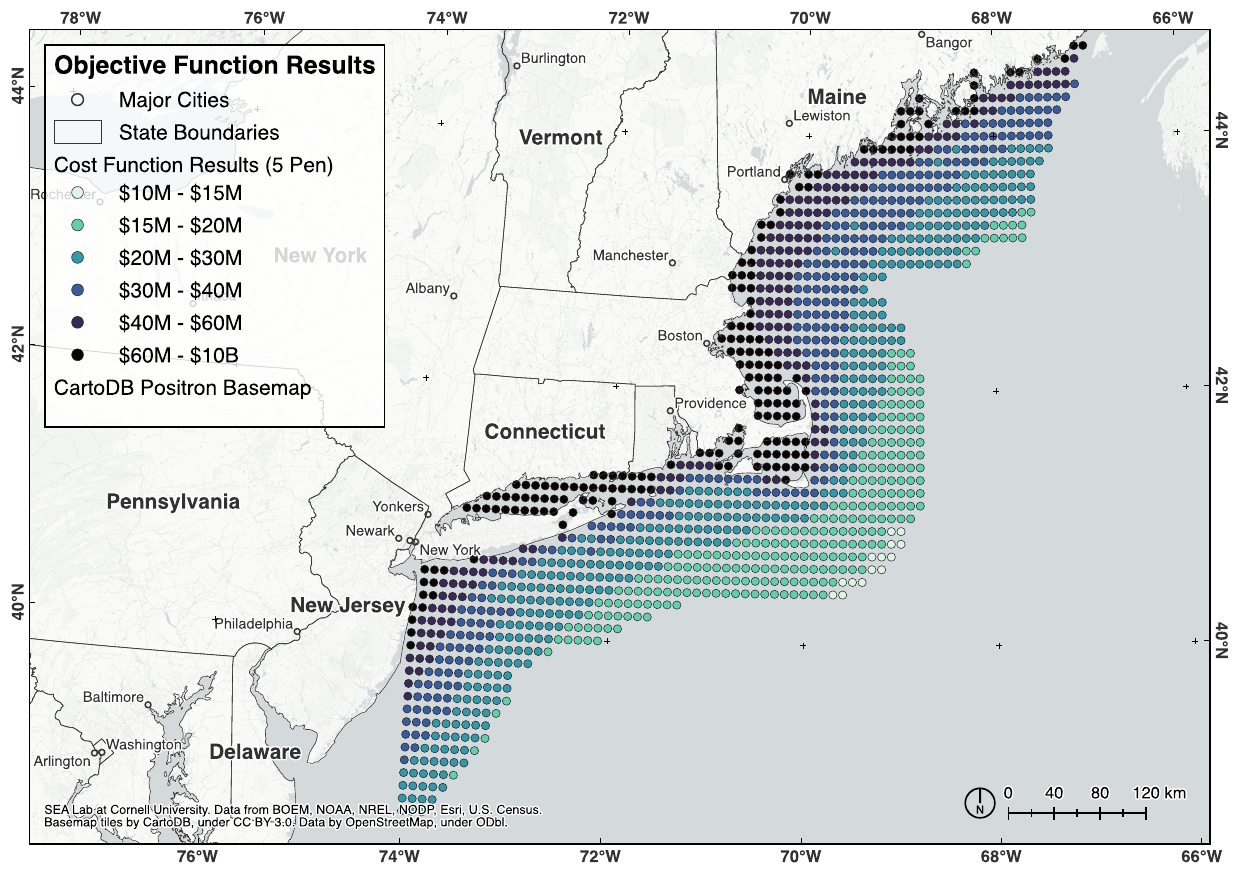}
    \centering
    \caption{Results for the cost function (i.e. location-dependent costs over the lifetime of the farm) on all points with data available. Location-dependent cost is inversely correlated with distance from shore.}
    \label{fig:complete-cost-results}
\end{figure}

In this section, the MSP results for a small WPAF with five net pens are presented. Although the conflict analysis significantly limits the number of feasible sites, a cost term is still initially calculated for each point in order to show overall trends, and the results are presented in figure \ref{fig:complete-cost-results}.  The results of the objective function calculating location-dependent costs across the region range from \$13.1M far from shore to \$4.0B in bays and sheltered areas near the coast where wave energy production is unrealistic.

\begin{figure}[t!]
    \includegraphics[width=\textwidth]{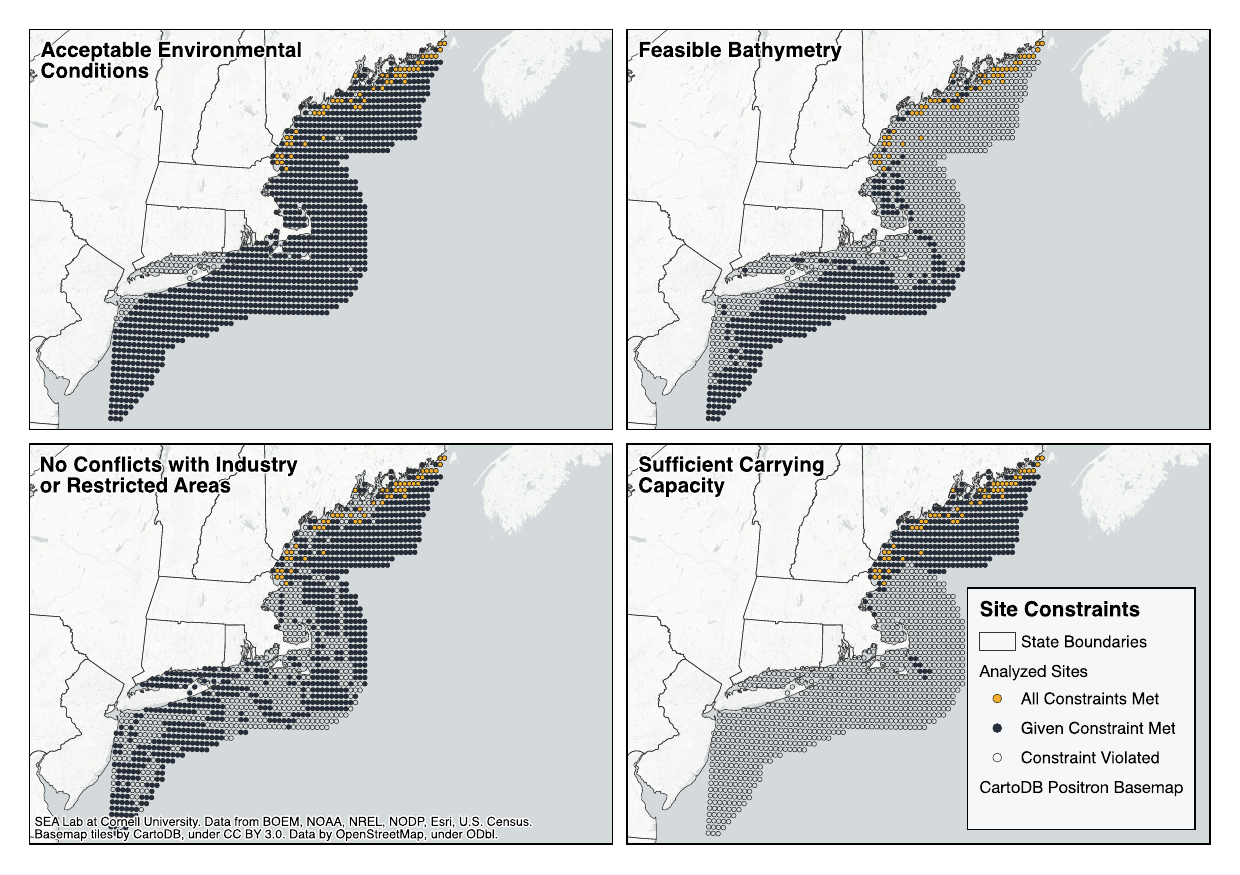}
    \centering
    \caption{Potential sites with data available are evaluated for constraints including acceptable environmental conditions, feasible bathymetry, no conflicts, and sufficient carrying capacity.}
    \label{fig:evaluated-points}
\end{figure}

\begin{figure}[t!]
    \includegraphics[width=\textwidth]{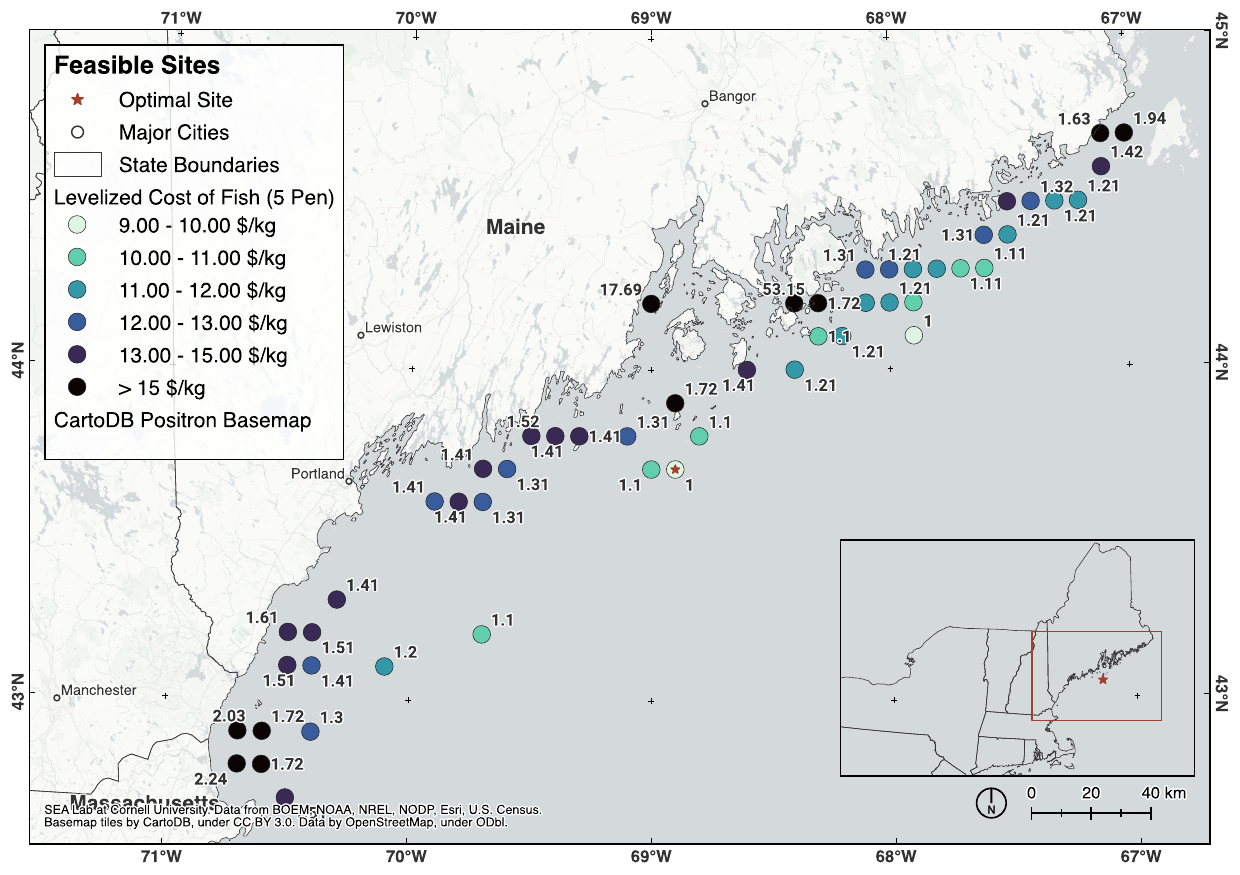}
    \centering
    \caption{Sites that meet requirements of depth, carrying capacity, and lack of conflicts. Colors indicate the levelized cost of fish, and numbers indicate the cost multiplier relative to the optimal site.}
    \label{fig:feasible-cost-results}
\end{figure}

Figure \ref{fig:evaluated-points} shows the results of imposing different constraints on the WPAF model. Given the resolution of {0.1\degree} latitude and longitude across this region, 7038 points are initially sampled, of which 3249 are offshore and only 1247 have data available for every condition in question. The number of feasible sites is further limited to 1182 feasible sites that meet acceptable environmental conditions, 529 sites with feasible bathymetry, 825 sites with no conflicts with industries or restricted areas, and 365 sites with sufficient carrying capacity. Together these constraints significantly narrow the feasible region and leave only 52 feasible sites ($\sim$4\%) along the coast of Maine that meet all the requirements, as shown in figure \ref{fig:feasible-cost-results}.

\begin{figure}[t!]
    \includegraphics[width=\textwidth]{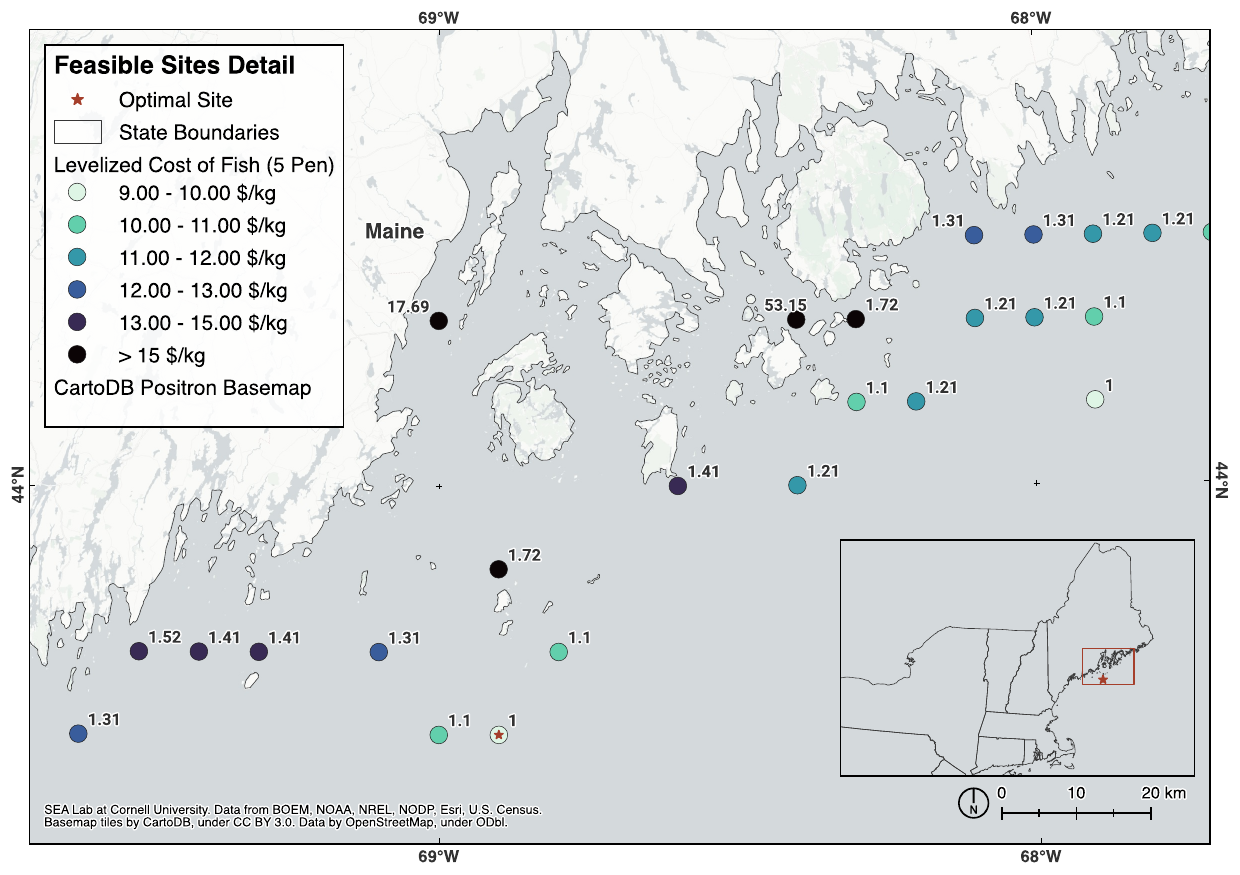}
    \centering
    \caption{Detail of optimal sites\ and nearby feasible sites. Colors indicate the levelized cost of fish, and numbers indicate the cost multiplier relative to the optimal site.}
    \label{fig:detail-cost-results}
\end{figure}

Combining both the objective function and constraints yields several final sites that meet the stated requirements, and one optimal site with the lowest cost. For feasible sites, the objective function ranges from \$30.0M to \$3.0B and follows a similar trend as the overall results. Feasible sites are shown in figure \ref{fig:feasible-cost-results}, with figure \ref{fig:detail-cost-results} providing further detail near the optimal site. Both figures show sites color-coded by their value for levelized cost of fish. The small numbers next to each site on the figure denote the levelized cost of fish as a multiple of the cost of the optimal site.

\begin{figure}[t!]
    \includegraphics[width=1\textwidth]{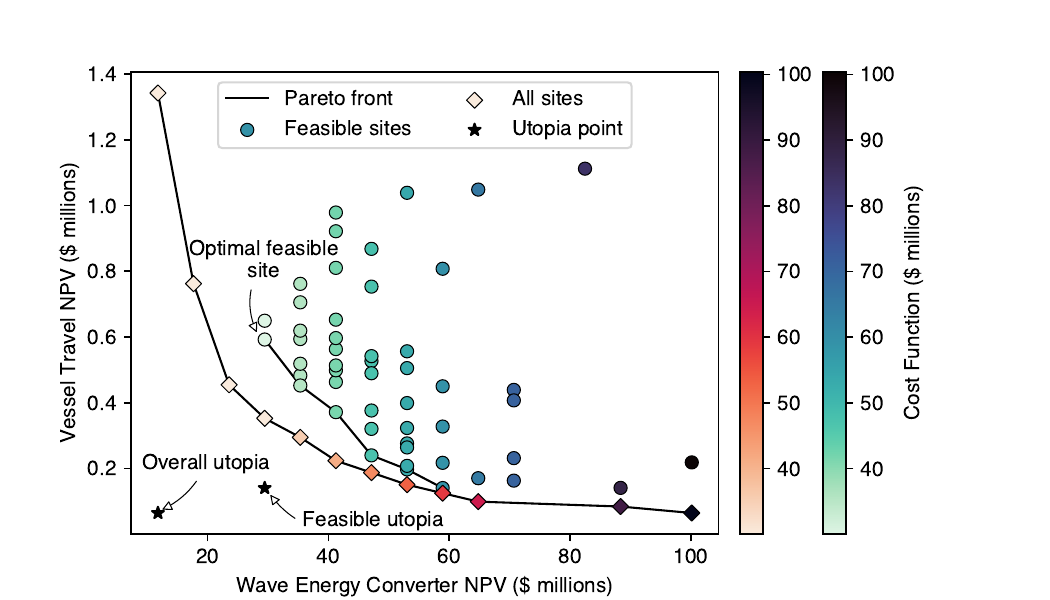}
    \centering
    \caption{Results of the vessel travel and WEC cost function for feasible sites (circles), and overall results regardless of whether they meet the constraints (diamonds). Note that sites with excessive costs (\$600M and above) are not shown on plot.}
    \label{fig:pareto-all}
\end{figure}

The scatter plot, shown in figure \ref{fig:pareto-all}, provides more detail on the results of the cost analysis and overall trends. The chart compares the effects of the two location-dependent cost terms in different subsets of the results. The diamond-shaped points show all non-dominated sites for the overall cost function without respect to constraints (figure \ref{fig:complete-cost-results}) while the circular points show all sites that meet the constraints (figure \ref{fig:feasible-cost-results}). For each of these sets, a Pareto front and utopia point is shown, and the overall cost function is indicated by the color gradient. The utopia point refers the hypothetical ideal point that exists at the optimal values for each objective. The figure shows the dominance of WEC cost in the overall cost function as the cheapest sites overall occur low on the WEC cost axis but vary only slightly with their position on the travel cost axis. The difference between feasible and overall non-dominated sites is also interesting. The cheapest feasible sites approach the overall Pareto front but fail to compete with the absolute cheapest sites because of the additional constraints.

An optimal site is identified southwest of Acadia National Park at 43.7\degree N 68.9\degree W. This site has location-dependent costs including WECs and vessel travel of \$30.0M and an overall cost of \$56.8M. A breakdown of all cost terms related to this site for a small WPAF with five cylindrical net pens is presented in table \ref{table:optimal-site-costs}. Note that these costs are reported for the WPAF considering a lifetime of 15 years, and expected annual fish harvest from the farm.

\begin{table}[!t]
\footnotesize
\caption{Overview of all cost terms for a sample small WPAF with five cylindrical net pens with a diameter of 30 m, height of 15 m, 15-year lifetime, and an annual fish yield of 676 tonnes at the optimal location of 43.7\degree N and 68.9\degree W (WEC cost accounts for CapEx and OpEx terms).}
\begin{center}
%\resizebox{\columnwidth}{!}{%
\begin{tabular}{| c | c | c |} 
 \hline
\textbf{Cost term} & \textbf{Value [\$ millions]}& \textbf{Share [\%]} \\
 \hline
WEC cost & 29.45 & 51.8 \\
 \hline
Vessel travel cost & 0.59 & 1.0 \\
\hline
Net pen cost & 5.30 & 9.3 \\
\hline
Feed barge cost & 3.73 & 6.6 \\
\hline
Fish feed cost & 12.8 & 22.5 \\
\hline
Fingerling cost & 4.94 & 8.7 \\
 \hline
Total cost & 56.81 & 100 \\
 \hline
\end{tabular}
%}
\end{center}
  \vspace{-5mm}
\label{table:optimal-site-costs}
\end{table}

It can be observed that WEC cost will account for more than half of the total cost of the WPAF, which highlights the importance of finding a location for this system with high wave power density and therefore a lower number of WECs. Using equation \ref{eq:cost-per-fish-yield} and considering the total cost and an annual fish yield of 676 tonnes, the levelized cost of fish at the optimal site is estimated to be \$9.23 per kilogram. This unit price is marginally higher ($\sim$11\%) than the current market rate which in 2023 placed salmon at \$8.28 per kilogram \citep{statista_salmon_2024}.

\section{Discussion}\label{sec:discussion}

Overall trends include a large amount of variation between the costs of each site despite relative proximity, driven in large part by access to wave resources. The decrease in cost further offshore demonstrates that location-dependent WPAF costs are dominated by the WEC cost, as opposed to the vessel travel cost which increases with distance from shore.  The optimal site at 43.7\degree N 68.9\degree W, with location-dependent costs of \$30.0M is one of the furthest feasible sites from shore and has location-dependent costs several times cheaper than sites closer to land. Closer to shore, the cost increases significantly for sites in sheltered bays and other areas that would require many more WECs to power even a modest farm. This trend by far outweighs factors like the increased travel costs for deployment locations further from shore, although there are certain limits due to increasing depth and other constraints. The main takeaway from this cost analysis is that it is worthwhile to consider deployment locations farther offshore to decrease the WPAF cost by decreasing the number of required WECs. At these locations, the model estimates a levelized cost of fish near current market rates. Although this may indicate that the model is within the ballpark of actual costs, the cost model currently lacks enough detail to draw further conclusions or to make comparisons with market rates. The cost model used in this study does not include all cost components and instead focuses on those that change with location, such as distance to shore. The result is a relative, rather than absolute, metric intended to enable a comparison of different sites within the same study.

The conflict analysis also reveals the difficulty in finding a feasible site for a WPAF, especially given the number of other oceanic activities and the number of areas that do not meet the required bathymetry or carrying capacity. The final results end up being further north than perhaps expected, far away from many of the areas where current activities like fishing and offshore wind development occur. In some cases, changes to the design or planning process may allow some of these constraints to be relaxed. Carrying capacity, for example, is the most limiting constraint in this case study and is strongly influenced by net pen size and fish stocking density. In this case study, these values are taken from current industrial net pen standards \citep{innovasea_innovasea_2023}, but decreasing the size of the net pen or its stocking density would increase the number of feasible points, albeit at the cost of decreased yield. Conflicts with certain offshore areas, like military zones, may never be feasible while others, like wind farms, may allow for co-location in the future as potentially synergistic relationships are better understood. Future work investigating this synergy could increase the number of feasible points for the WPAF given the possibility of co-locating the wind farm and WPAF at the same site.

Existing WPAF studies demonstrate similar methodologies and trends, although it is difficult to compare results directly given that WPAFs remain a relatively new concept. A study by \citet{clemente_wave_2023} investigates co-location potential at two sites along the Portuguese coastline using various farm designs. Although \citet{clemente_wave_2023} do consider some similar factors including annual wave energy availability, a point absorber WEC as an energy source, and the feasibility of Atlantic salmon, their analysis is more focused on changing the design of the farm at a limited number of sites rather than analyzing many different sites with a single farm design. A study by \citet{garavelli_feasibility_2022} is more similar to this study in that it focuses on siting and considers similar geographic factors including bathymetry, wave height, and marine protected areas. Their analysis yields similar results to this study as they identify wave power density, bathymetry, and conflicts (in their case navigation routes and managed areas) to be key factors in determining favorable locations. However, they use a binary suitability score rather than a cost model making it difficult to directly compare the impact of these criteria on site selection.

\section{Conclusion}

Marine spatial planning is a topic of critical importance given the continued development of various offshore industries and our increasing reliance on aquatic resources for food and energy. Careful planning of these resources is necessary to ensure their sustainable use and avoid conflicts with other existing and future industries that operate along our coastlines. This paper provided an overview of several MSP techniques using open-source software including QGIS and Python, alongside the application of those techniques to a case study in siting a wave-powered aquaculture farm.

The MSP techniques discussed include the process of selecting a scope and preparing data, analysis options within QGIS of both vector and raster data sets as they relate to marine energy applications, and the potential to integrate with more complex Python models using GeoPandas, Rasterio, and a ``handler'' object made for the model used in the case study. Using information on the wave climate, biological conditions, and various conflicts, several feasible sites for a wave-powered aquaculture farm were identified along the coast of Maine. A cost model was then used to select one optimal location southwest of Acadia National Park with location-dependent costs of \$30.0M for a five-pen farm. The optimal site has a levelized cost of fish of \$9.23 per kilogram, which is on par with current market rates. The analysis revealed the importance of identifying conflicts in the siting process early on, and the significance of wave energy availability, and therefore distance to shore, to the cost of a project.

Future work on the case study of a wave-powered aquaculture farm includes more in-depth analysis given time series data of conditions and consideration of energy storage within the WPAF to ensure a reliable power supply. Different layouts, directions, near and far-field effects, and park effects could also be integrated to provide more detailed modeling of WEC performance. The cost model could also be improved by adding terms related to the energy transfer between the WEC and aquaculture farm. An integrated multi-trophic aquaculture setup will also be considered which allows other species such as kelp and mollusks to grow alongside fish with the goal of creating mutualistic relationships between the different species. This could be coupled with an expansion of the carrying capacity model to include aspects like nutrient use and production in addition to oxygen depletion. Additional environmental metrics would be useful to integrate as well, including constraints on benthic habitat and areas with endangered species, and emissions savings over a diesel-powered farm. The MSP techniques and generated map layers from this study will also be integrated into a more general web-based mapping tool to make the work more accessible to the many stakeholders who could benefit from this kind of analysis. Finally, more outreach with stakeholders will be conducted to receive feedback on the usefulness of the existing analysis and more information on the issues they face in practice.

\section*{Acknowledgments}
This work was supported in part by Sea Grant Regional Research Project No. R/ATD-18-NESG and the Engineering Learning Initiatives program at Cornell University. We would also like to thank Rebecca McCabe and Matthew Haefner; members of the Symbiotic Engineering and Analysis (SEA) Lab; and our industry collaborators CalWave Power Technologies, Innovasea, and Manna Fish Farms, for providing feedback and support throughout this study.

\section*{Data Availability}
Code required to reproduce the findings in this study is available in the SEA Lab GitHub at \href{https://github.com/symbiotic-engineering/aquaculture}{https://github.com/symbiotic-engineering/aquaculture}, and the relevant pre-processed data files are hosted on Zenodo at \href{https://zenodo.org/records/10140826}{https://zenodo.org/records/10140826}. Raw data files are publicly available from the sources cited in tables \ref{table:data-sources-raster} and \ref{table:data-sources-vector}.

\appendix
\section{QGIS Instructions}
\label{appendix:qgis-instructions}

This appendix contains a brief overview of specific QGIS instructions for the steps outlined in this paper for those interested in following a similar procedure for MSP analysis. The QGIS Training Manual (\href{https://docs.qgis.org/latest/en/docs/training_manual/index.html}{https://docs.qgis.org/latest/en/docs/training\_manual/index.html}) and associated documentation are also helpful guides which provide more detail and further instructions. The processes below correspond to and are further explained in sections \ref{sec:preprocessing-data}, \ref{sec:raster-analysis}, and \ref{sec:vector-analysis}.

\subsection{File Conversion}
Conversion from an ArcGIS format, is only necessary for certain raster files distributed in the proprietary GeoDatabase (.gdb) format. Although QGIS can currently access vector data in these files, rasters must be loaded into ArcGIS and exported as GeoTIFF (.tiff) files to be readable by QGIS and other applications. This process is straightforward but unfortunately requires users interested in using data from organizations that publish data in GeoDatabases to have access to ArcGIS.

\subsection{Vector interpolation}

Certain data sets for environmental conditions are distributed in a vector format with many points containing measurements instead of as a continuous raster. For many analyses, having a continuous raster of conditions is beneficial, and interpolation can be used to approximate this from discrete measurements. In this study, the data used for dissolved oxygen and salinity are distributed as vector files, but are converted to rasters. From the processing toolbox (\texttt{Processing > Toolbox}), the \texttt{TIN Interpolation} tool can be used to convert interpolate a vector data set of point measurements into a continuous raster data set. \texttt{TIN Interpolation} allows the user to select the desired layer, the attribute that contains the desired measurement, and a reasonable resolution (as discussed in section \ref{subsec:study-scope}) for the given study. The resulting raster is limited to the area bounded by the original points, so \texttt{r.fill.stats} in the processing toolbox can be used to extend the range of the raster and smooth the results of the interpolation. The parameters of this operation will vary by the study scope and individual data set. For this study, each raster is extended past the shoreline to ensure all coastal waters could be analyzed, then clipped back to include only the study scope.

\subsection{Clipping}
Data is clipped to only include the scope of designated state and federal waters. This relies on a vector layer describing the study scope which, in this study, is the file with state and federal waters from NOAA in table 2. To clip raster layers, click \texttt{Raster > Extraction > Clip Raster by Mask Layer...} and select the correct vector layer as a mask.  For vector layers, click \texttt{Vector > Geoprocessing Tools > Clip...} and select the same mask as the overlay layer. Both operations will create a temporary scratch layer of the results, which can be exported to a permanent file.

\subsection{Re-projection}
GIS data and maps use different projections to translate the three-dimensional globe into a two-dimensional image. This study uses WGS 84 (EPSG:4326), a very common CRS, to standardize all downloaded data because of its simplicity, and so that each data set uses units of degrees in latitude and longitude. WGS 84 / UTM Zone 19 is used for final maps exported from QGIS, which is tailored for the specific region of the earth included in the scope and minimizes distortion in that area. Layers can be re-projected during the export process. Simply right-click and select \texttt{Export > Save Feature As...}, then select the desired projection (e.g. WGS 84) as the CRS. The CRS for final maps (e.g. WGS 84 / UTM Zone 19) is set later in the item properties of a map element in a layout, and does not need to match the CRS of individual layers.

\subsection{Contour generation}
Contours are generated via \texttt{Raster > Extraction > Contour...}. An input layer can be selected alongside an interval size between each line, which should be picked based on the range of values in the particular data set. The attribute name, which defaults to \texttt{ELEV}, is the name of the field in the contour data set with measurement values. After generating contours, it can be helpful to style them or add labels in \texttt{Layer Styling > Labels} or in some cases filter out some the contours that are shown on the map. In the case of bathymetry, for example, the zero-value contour is filtered out along with many of the deeper regions where contours become more tightly spaced along the continental shelf.

\subsection{Reclassification}
The Raster Calculator (\texttt{Raster > Raster Calculator...}) is useful for a many operations including reclassification of raster data sets. To reclassify about the mean as in the case of fishing traffic, identify the mean and standard deviation in the layer properties dialog. The code below is an example used to reclassify fishing data given a mean of 4.6 vessels and standard deviation of 23.1.
\begin{verbatim}
("2021 All Vessel Activity@1" - 4.6) / 23.1
\end{verbatim}
This generates a new output layer with fishing data centered around the mean which can be easily displayed or converted to a vector file as in the case of a conflicts analysis. Raster Calculator tends to output somewhat messy files, but these problems can be addressed with some additional clipping and layer styling.

\subsection{Polygonalization}
\texttt{Raster > Conversion > Polygonize...} can be run given the input raster and band (the component of a layer with a specific measurement) to convert the raster to vector areas. Depending on the input data file, the resulting vector will likely appear rough and may contain island-like areas that are too small to be relevant for analysis on a larger scale. There are several processing tools to improve the generated vector including \texttt{Delete holes}, which can fill in small areas; \texttt{Add geometry attributes}, which can calculate the area of each region thus allowing very small regions to be filtered or deleted; and \texttt{Buffer}, which can be used to expand or contract regions and create a smoother outer edge than what is generated from the original raster pixels.

\subsection{Intersection identification}
To find the intersection between two vector layers, use \texttt{Vector > GeoProcessing Tools > Intersection} and select the layers in question. This generates a new layer of the intersection which can be stylized to highlight conflicts or other areas of importance.

\Urlmuskip=0mu plus 1mu
\bibliographystyle{unsrtnat}
\bibliography{msp-refs.bib}

\begin{thebibliography}{48}
\providecommand{\natexlab}[1]{#1}
\providecommand{\url}[1]{\texttt{#1}}
\expandafter\ifx\csname urlstyle\endcsname\relax
  \providecommand{\doi}[1]{doi: #1}\else
  \providecommand{\doi}{doi: \begingroup \urlstyle{rm}\Url}\fi

\bibitem[Arbo and Th{\h u}y(2016)]{arbo_use_2016}
Peter Arbo and Ph\d am Th\d i~Thanh Th{\h u}y.
\newblock Use conflicts in marine ecosystem-based management - {The} case of
  oil versus fisheries.
\newblock \emph{Ocean \& Coastal Management}, 122:\penalty0 77--86, March 2016.
\newblock \doi{10.1016/j.ocecoaman.2016.01.008}.
\newblock URL
  \url{https://www.sciencedirect.com/science/article/pii/S0964569116300084}.

\bibitem[Andrews et~al.(2021)Andrews, Bennett, Le~Billon, Green,
  Cisneros-Montemayor, Amongin, Gray, and Sumaila]{andrews_oil_2021}
Nathan Andrews, Nathan~J. Bennett, Philippe Le~Billon, Stephanie~J. Green,
  Andr\'es~M. Cisneros-Montemayor, Sandra Amongin, Noella~J. Gray, and
  U.~Rashid Sumaila.
\newblock Oil, fisheries and coastal communities: {A} review of impacts on the
  environment, livelihoods, space and governance.
\newblock \emph{Energy Research \& Social Science}, 75:\penalty0 102009, May
  2021.
\newblock \doi{10.1016/j.erss.2021.102009}.
\newblock URL
  \url{https://www.sciencedirect.com/science/article/pii/S221462962100102X}.

\bibitem[Bess and Rallapudi(2007)]{bess_spatial_2007}
Randall Bess and Ramana Rallapudi.
\newblock Spatial conflicts in {New} {Zealand} fisheries: {The} rights of
  fishers and protection of the marine environment.
\newblock \emph{Marine Policy}, 31\penalty0 (6):\penalty0 719--729, November
  2007.
\newblock \doi{10.1016/j.marpol.2006.12.009}.
\newblock URL
  \url{https://www.sciencedirect.com/science/article/pii/S0308597X07000152}.

\bibitem[{NOAA Fisheries}(2023)]{noaa_fisheries_offshore_2023}
{NOAA Fisheries}.
\newblock Offshore {Aquaculture} and the {Future} of {Sustainable} {Seafood}
  {\textbar} {NOAA} {Fisheries}.
\newblock \emph{NOAA}, October 2023.
\newblock URL
  \url{https://www.fisheries.noaa.gov/feature-story/offshore-aquaculture-and-future-sustainable-seafood}.
\newblock Archive Location: Southeast.

\bibitem[Gentry et~al.(2017)Gentry, Froehlich, Grimm, Kareiva, Parke, Rust,
  Gaines, and Halpern]{gentry_mapping_2017}
Rebecca~R. Gentry, Halley~E. Froehlich, Dietmar Grimm, Peter Kareiva, Michael
  Parke, Michael Rust, Steven~D. Gaines, and Benjamin~S. Halpern.
\newblock Mapping the global potential for marine aquaculture.
\newblock \emph{Nature Ecology \& Evolution}, 1\penalty0 (9):\penalty0
  1317--1324, September 2017.
\newblock \doi{10.1038/s41559-017-0257-9}.
\newblock URL \url{https://www.nature.com/articles/s41559-017-0257-9}.

\bibitem[Weiss et~al.(2018)Weiss, Ondiviela, Guanche, Castellanos, and
  Juanes]{weiss_global_2018}
Carlos V.~C. Weiss, B\'arbara Ondiviela, Ra\'ul Guanche, Omar~F. Castellanos,
  and Jos\'e~A. Juanes.
\newblock A global integrated analysis of open sea fish farming opportunities.
\newblock \emph{Aquaculture}, 497:\penalty0 234--245, December 2018.
\newblock \doi{10.1016/j.aquaculture.2018.07.054}.
\newblock URL
  \url{https://www.sciencedirect.com/science/article/pii/S0044848618304605}.

\bibitem[Brugere(2006)]{brugere_can_2006}
Cecile Brugere.
\newblock Can integrated coastal management solve
  agriculture-fisheries-aquaculture conflicts at the land-water interface? a
  perspective from new institutional economics.
\newblock \emph{Environment and Livelihoods in Tropical Coastal Zones: Managing
  Agriculture-Fishery-Aquaculture Conflicts}, July 2006.
\newblock \doi{10.1079/9781845931070.0258}.
\newblock URL
  \url{https://www.cabidigitallibrary.org/doi/10.1079/9781845931070.0258}.

\bibitem[Bishwajit(2014)]{bishwajit_fisheries_2014}
Ghose Bishwajit.
\newblock Fisheries and {Aquaculture} in {Bangladesh}: {Challenges} and
  {Opportunities}.
\newblock \emph{Annals of Aquaculture and Research}, 1, July 2014.
\newblock URL
  \url{https://www.researchgate.net/publication/316585813_Fisheries_and_Aquaculture_in_Bangladesh_Challenges_and_Opportunities}.

\bibitem[Longdill et~al.(2008)Longdill, Healy, and
  Black]{longdill_integrated_2008}
Peter~C. Longdill, Terry~R. Healy, and Kerry~P. Black.
\newblock An integrated {GIS} approach for sustainable aquaculture management
  area site selection.
\newblock \emph{Ocean \& Coastal Management}, 51\penalty0 (8):\penalty0
  612--624, January 2008.
\newblock \doi{10.1016/j.ocecoaman.2008.06.010}.
\newblock URL
  \url{https://www.sciencedirect.com/science/article/pii/S0964569108000604}.

\bibitem[Stelzenm\"uller et~al.(2022)Stelzenm\"uller, Letschert, Gimpel, Kraan,
  Probst, Degraer, and D\"oring]{stelzenmuller_plate_2022}
V.~Stelzenm\"uller, J.~Letschert, A.~Gimpel, C.~Kraan, W.~N. Probst,
  S.~Degraer, and R.~D\"oring.
\newblock From plate to plug: {The} impact of offshore renewables on {European}
  fisheries and the role of marine spatial planning.
\newblock \emph{Renewable and Sustainable Energy Reviews}, 158:\penalty0
  112108, April 2022.
\newblock \doi{10.1016/j.rser.2022.112108}.
\newblock URL
  \url{https://www.sciencedirect.com/science/article/pii/S1364032122000375}.

\bibitem[Firestone et~al.(2020)Firestone, Hirt, Bidwell, Gardner, and
  Dwyer]{firestone_faring_2020}
Jeremy Firestone, Christine Hirt, David Bidwell, Meryl Gardner, and Joseph
  Dwyer.
\newblock Faring well in offshore wind power siting? {Trust}, engagement and
  process fairness in the {United} {States}.
\newblock \emph{Energy Research \& Social Science}, 62:\penalty0 101393, April
  2020.
\newblock \doi{10.1016/j.erss.2019.101393}.
\newblock URL
  \url{https://www.sciencedirect.com/science/article/pii/S2214629619306553}.

\bibitem[ten Brink and Dalton(2018)]{ten_brink_perceptions_2018}
Talya ten Brink and Tracey Dalton.
\newblock Perceptions of {Commercial} and {Recreational} {Fishers} on the
  {Potential} {Ecological} {Impacts} of the {Block} {Island} {Wind} {Farm}
  ({US}).
\newblock \emph{Frontiers in Marine Science}, 5:\penalty0 439, November 2018.
\newblock \doi{10.3389/fmars.2018.00439}.
\newblock URL
  \url{https://www.frontiersin.org/articles/10.3389/fmars.2018.00439/full}.

\bibitem[Buck et~al.(2004)Buck, Krause, and Rosenthal]{buck_extensive_2004}
Bela~Hieronymus Buck, Gesche Krause, and Harald Rosenthal.
\newblock Extensive open ocean aquaculture development within wind farms in
  {Germany}: the prospect of offshore co-management and legal constraints.
\newblock \emph{Ocean \& Coastal Management}, 47\penalty0 (3):\penalty0
  95--122, January 2004.
\newblock \doi{10.1016/j.ocecoaman.2004.04.002}.
\newblock URL
  \url{https://www.sciencedirect.com/science/article/pii/S0964569104000262}.

\bibitem[Buck et~al.(2008)Buck, Krause, Michler-Cieluch, Brenner, Buchholz,
  Busch, Fisch, Geisen, and Zielinski]{buck_meeting_2008}
B.~H. Buck, G.~Krause, T.~Michler-Cieluch, M.~Brenner, C.~M. Buchholz, J.~A.
  Busch, R.~Fisch, M.~Geisen, and O.~Zielinski.
\newblock Meeting the quest for spatial efficiency: progress and prospects of
  extensive aquaculture within offshore wind farms.
\newblock \emph{Helgoland Marine Research}, 62\penalty0 (3):\penalty0 269--281,
  September 2008.
\newblock \doi{10.1007/s10152-008-0115-x}.
\newblock URL
  \url{https://hmr.biomedcentral.com/articles/10.1007/s10152-008-0115-x}.

\bibitem[Gimpel et~al.(2015)Gimpel, Stelzenm\"uller, Grote, Buck, Floeter,
  N\'u\~nez Riboni, Pogoda, and Temming]{gimpel_gis_2015}
Antje Gimpel, Vanessa Stelzenm\"uller, Britta Grote, Bela~H. Buck, Jens
  Floeter, Ismael N\'u\~nez Riboni, Bernadette Pogoda, and Axel Temming.
\newblock A {GIS} modelling framework to evaluate marine spatial planning
  scenarios: {Co}-location of offshore wind farms and aquaculture in the
  {German} {EEZ}.
\newblock \emph{Marine Policy}, 55:\penalty0 102--115, May 2015.
\newblock \doi{10.1016/j.marpol.2015.01.012}.
\newblock URL
  \url{https://www.sciencedirect.com/science/article/pii/S0308597X15000238}.

\bibitem[Wever et~al.(2015)Wever, Krause, and Buck]{wever_lessons_2015}
Lara Wever, Gesche Krause, and Bela~H. Buck.
\newblock Lessons from stakeholder dialogues on marine aquaculture in offshore
  wind farms: {Perceived} potentials, constraints and research gaps.
\newblock \emph{Marine Policy}, 51:\penalty0 251--259, January 2015.
\newblock \doi{10.1016/j.marpol.2014.08.015}.
\newblock URL
  \url{https://www.sciencedirect.com/science/article/pii/S0308597X14002310}.

\bibitem[Garavelli et~al.(2022)Garavelli, Freeman, Tugade, Greene, and
  McNally]{garavelli_feasibility_2022}
Lysel Garavelli, Mikaela~C. Freeman, Levy~G. Tugade, David Greene, and Jim
  McNally.
\newblock A feasibility assessment for co-locating and powering offshore
  aquaculture with wave energy in the {United} {States}.
\newblock \emph{Ocean \& Coastal Management}, 225:\penalty0 106242, June 2022.
\newblock \doi{10.1016/j.ocecoaman.2022.106242}.
\newblock URL
  \url{https://www.sciencedirect.com/science/article/pii/S0964569122002186}.

\bibitem[Clemente et~al.(2023)Clemente, Rosa-Santos, Ferradosa, and
  Taveira-Pinto]{clemente_wave_2023}
D.~Clemente, P.~Rosa-Santos, T.~Ferradosa, and F.~Taveira-Pinto.
\newblock Wave energy conversion energizing offshore aquaculture: {Prospects}
  along the {Portuguese} coastline.
\newblock \emph{Renewable Energy}, 204:\penalty0 347--358, March 2023.
\newblock \doi{10.1016/j.renene.2023.01.009}.
\newblock URL
  \url{https://www.sciencedirect.com/science/article/pii/S0960148123000095}.

\bibitem[Whiting et~al.(2023)Whiting, Garavelli, Farr, and
  Copping]{whiting_effects_2023}
Jonathan Whiting, Lysel Garavelli, Hayley Farr, and Andrea Copping.
\newblock Effects of small marine energy deployments on oceanographic systems.
\newblock \emph{International Marine Energy Journal}, 6\penalty0 (2):\penalty0
  45--54, December 2023.
\newblock \doi{10.36688/imej.6.45-54}.
\newblock URL \url{https://marineenergyjournal.org/imej/article/view/100}.

\bibitem[Silva et~al.(2018)Silva, Rusu, and Guedes~Soares]{silva_effect_2018}
Dina Silva, Eugen Rusu, and C.~Guedes~Soares.
\newblock The {Effect} of a {Wave} {Energy} {Farm} {Protecting} an
  {Aquaculture} {Installation}.
\newblock \emph{Energies}, 11\penalty0 (8):\penalty0 2109, August 2018.
\newblock \doi{10.3390/en11082109}.
\newblock URL \url{https://www.mdpi.com/1996-1073/11/8/2109}.

\bibitem[Hasankhani et~al.(2023{\natexlab{a}})Hasankhani, McCabe, Ewig, Won,
  and Haji]{hasankhani_conceptual_2023}
Arezoo Hasankhani, Rebecca McCabe, Gabriel Ewig, Eugene~Thome Won, and Maha~N
  Haji.
\newblock Conceptual {Design} and {Optimization} of a {Wave}-{Powered}
  {Offshore} {Aquaculture} {Farm}.
\newblock In \emph{The 33rd {International} {Ocean} and {Polar} {Engineering}
  {Conference}}, Ottawa, Canada, June 2023{\natexlab{a}}.
\newblock URL
  \url{https://onepetro.org/ISOPEIOPEC/proceedings/ISOPE23/All-ISOPE23/ISOPE-I-23-112/524501}.

\bibitem[Hasankhani et~al.(2023{\natexlab{b}})Hasankhani, Ewig, McCabe, Won,
  and Haji]{hasankhani_marine_2023}
Arezoo Hasankhani, Gabriel Ewig, Rebecca McCabe, Eugene~Thome Won, and Maha~N
  Haji.
\newblock Marine {Spatial} {Planning} of a {Wave}-{Powered} {Offshore}
  {Aquaculture} {Farm} in the {Northeast} {U}.{S}.
\newblock In \emph{{OCEANS} 2023 - {Limerick}}, pages 1--10, Limerick, Ireland,
  June 2023{\natexlab{b}}.
\newblock \doi{10.1109/OCEANSLimerick52467.2023.10244332}.
\newblock URL \url{https://ieeexplore.ieee.org/document/10244332}.

\bibitem[Neary et~al.(2014)Neary, Previsic, Jepsen, Lawson, Yu, Copping,
  Fontaine, Hallett, and Murray]{neary_methodology_2014}
Vincent~S Neary, Mirko Previsic, Richard~A Jepsen, Michael~J Lawson, Yi-Hsiang
  Yu, Andrea~E Copping, Arnold~A Fontaine, Kathleen~C Hallett, and Dianne~K
  Murray.
\newblock Methodology for {Design} and {Economic} {Analysis} of {Marine}
  {Energy} {Conversion} ({MEC}) {Technologies}.
\newblock \emph{Sandia National Laboratories}, page 262, March 2014.
\newblock URL
  \url{https://energy.sandia.gov/wp-content/gallery/uploads/SAND2014-9040-RMP-REPORT.pdf}.

\bibitem[Zhong and Yeung(2019)]{zhong_wave-body_2019}
Qian Zhong and Ronald~W. Yeung.
\newblock Wave-body interactions among energy absorbers in a wave farm.
\newblock \emph{Applied Energy}, 233-234:\penalty0 1051--1064, January 2019.
\newblock \doi{10.1016/j.apenergy.2018.09.131}.
\newblock URL
  \url{https://www.sciencedirect.com/science/article/pii/S0306261918314442}.

\bibitem[{Scale Aquaculture AS}(2019)]{scale_aquaculture_as_seafarm_2019}
{Scale Aquaculture AS}.
\newblock {SeaFarm} {Feeder} {Bow} 200, 2019.
\newblock URL \url{https://scaleaq.com/product/seafarm-feeder-bow-200/}.

\bibitem[{Innovasea}(2023)]{innovasea_innovasea_2023}
{Innovasea}.
\newblock Innovasea, 2023.
\newblock URL \url{https://www.innovasea.com/open-ocean-aquaculture/}.

\bibitem[Whittle(2022)]{whittle_haul_2022}
Patrick Whittle.
\newblock Haul of {Atlantic} cod, once abundant, reaches new low.
\newblock \emph{WBUR}, March 2022.
\newblock URL
  \url{https://www.wbur.org/news/2022/05/10/maine-massachusetts-cod-fishing-industry-record-low-catch}.

\bibitem[{Northeast Sea Grant}(2022)]{northeast_sea_grant_northeast_2022}
{Northeast Sea Grant}.
\newblock Northeast {Sea} {Grant} {Consortium}, August 2022.
\newblock URL \url{https://www.northeastseagrant.com}.

\bibitem[{Northeast Ocean Data Portal} and
  Shmookler(2016)]{northeast_ocean_data_portal_fvcom_2016}
{Northeast Ocean Data Portal} and Rachel Shmookler.
\newblock {FVCOM} {Annual} {Climatology} for {Temperature}, {Stratification},
  and {Currents}, February 2016.
\newblock URL
  \url{https://www.northeastoceandata.org/files/metadata/Themes/Habitat/FVCOMAnnualClimatology.pdf}.

\bibitem[{National Centers for Environmental
  Information}(2019)]{national_centers_for_environmental_information_world_2019}
{National Centers for Environmental Information}.
\newblock World {Ocean} {Atlas} 2018 {Data} {Access}, July 2019.
\newblock URL \url{https://www.ncei.noaa.gov/access/world-ocean-atlas-2018/}.

\bibitem[{National Renewable Energy
  Laboratory}(2011)]{national_renewable_energy_laboratory_marine_2011}
{National Renewable Energy Laboratory}.
\newblock Marine and {Hydrokinetic} {Resource} {Maps} and {Data}, October 2011.
\newblock URL \url{https://www.nrel.gov/gis/maps-marine.html}.

\bibitem[{National Geophysical Data
  Center}(1990)]{national_geophysical_data_center_bathymetry_1990}
{National Geophysical Data Center}.
\newblock Bathymetry, January 1990.
\newblock URL \url{https://www.northeastoceandata.org/data-download/}.

\bibitem[{Office for Coastal
  Management}(2019)]{office_for_coastal_management_principal_2019}
{Office for Coastal Management}.
\newblock Principal {Ports}, May 2019.
\newblock URL \url{https://www.fisheries.noaa.gov/inport/item/56124}.

\bibitem[Fontenault(2022)]{northeast_ocean_data_portal_fishing_2022}
Jeremy Fontenault.
\newblock Fishing {Vessel} {Transit} {Counts} from - 2021 {AIS}, May 2022.
\newblock URL
  \url{https://www.northeastoceandata.org/files/metadata/Themes/AIS/FishingAISVesselTransitCounts2021.pdf}.

\bibitem[{National Marine Protected Areas
  Center}(2020)]{national_marine_protected_areas_center_mpa_2020}
{National Marine Protected Areas Center}.
\newblock The {MPA} {Inventory}, 2020.
\newblock URL
  \url{https://marineprotectedareas.noaa.gov/dataanalysis/mpainventory/}.

\bibitem[{Northeast Ocean Data Portal} et~al.(2016){Northeast Ocean Data
  Portal}, {Naval Facilities Engineering Command (NAVFAC) Atlantic}, and
  {Ecology and Environment, Inc.}]{northeast_ocean_data_portal_national_2016}
{Northeast Ocean Data Portal}, {Naval Facilities Engineering Command (NAVFAC)
  Atlantic}, and {Ecology and Environment, Inc.}
\newblock National {Security} {Layers}, 2016.
\newblock URL \url{https://www.northeastoceandata.org/data-download/}.

\bibitem[{Office for Coastal
  Management}(2022)]{office_for_coastal_management_danger_2022}
{Office for Coastal Management}.
\newblock Danger {Zones} and {Restricted} {Areas}, October 2022.
\newblock URL \url{https://www.fisheries.noaa.gov/inport/item/48876}.

\bibitem[{Northeast Ocean Data
  Portal}(2015)]{northeast_ocean_data_portal_wind_2015}
{Northeast Ocean Data Portal}.
\newblock Wind {Energy} {Areas}, 2015 {Massachusetts} {Ocean} {Management}
  {Plan}, January 2015.
\newblock URL
  \url{https://www.northeastoceandata.org/files/metadata/Themes/EnergyAndInfrastructure/moris_om_wind_energy_areas_poly.htm}.

\bibitem[{Bureau of Ocean Energy
  Management}(2023)]{bureau_of_ocean_energy_management_renewable_2023}
{Bureau of Ocean Energy Management}.
\newblock Renewable {Energy} {Leases} and {Planning} {Areas}, 2023.
\newblock URL
  \url{https://www.boem.gov/renewable-energy/mapping-and-data/renewable-energy-gis-data}.

\bibitem[{Northeast Ocean Data
  Portal}(2010)]{northeast_ocean_data_portal_rhode_2010}
{Northeast Ocean Data Portal}.
\newblock Rhode {Island} {Renewable} {Energy} {Zone}, August 2010.
\newblock URL
  \url{https://www.northeastoceandata.org/files/metadata/Themes/EnergyAndInfrastructure/RenewableEnergyZone.htm}.

\bibitem[{Office of Coast Survey}(2015)]{office_of_coast_survey_shipping_2015}
{Office of Coast Survey}.
\newblock Shipping {Fairways}, {Lanes}, and {Zones} for {US} waters from
  2010-06-15 to 2010-08-15, December 2015.
\newblock URL \url{https://www.fisheries.noaa.gov/inport/item/39986}.

\bibitem[{Office for Coastal
  Management}(2018)]{office_for_coastal_management_federal_2018}
{Office for Coastal Management}.
\newblock Federal and {State} {Waters} from 2010-06-15 to 2010-08-15, April
  2018.
\newblock URL \url{https://www.fisheries.noaa.gov/inport/item/54383}.

\bibitem[{ESRI}(2023)]{esri_usa_2023}
{ESRI}.
\newblock {USA} {Major} {Cities}, 2023.
\newblock URL
  \url{https://hub.arcgis.com/datasets/esri::usa-major-cities/explore}.

\bibitem[{FEMP}(2021)]{femp_2021_2021}
{FEMP}.
\newblock 2021 {Discount} {Rates}.
\newblock Technical report, DOE, April 2021.
\newblock URL \url{https://www.energy.gov/femp/articles/2021-discount-rates}.

\bibitem[Freeman et~al.(2022)Freeman, Garavelli, Wilson, Hemer, Abundo, and
  Travis]{freeman_offshore_2022}
Mikaela~C. Freeman, Lysel Garavelli, Eloise Wilson, Mark Hemer, Michael
  Lochinvar~Sim Abundo, and Leonard~Edward Travis.
\newblock Offshore {Aquaculture}: {A} {Market} for {Ocean} {Renewable}
  {Energy}.
\newblock Technical report, Implementing Agreement on Ocean Energy System,
  April 2022.
\newblock URL
  \url{https://www.ocean-energy-systems.org/publications/oes-documents/market-policy-/document/offshore-aquaculture-a-market-for-ocean-renewable-energy./}.

\bibitem[Stigebrandt et~al.(2004)Stigebrandt, Aure, Ervik, and
  Hansen]{stigebrandt_regulating_2004}
Anders Stigebrandt, Jan Aure, Arne Ervik, and Pia~Kupka Hansen.
\newblock Regulating the local environmental impact of intensive marine fish
  farming: {III}. {A} model for estimation of the holding capacity in the
  {Modelling}-{Ongrowing} fish farm-{Monitoring} system.
\newblock \emph{Aquaculture}, 234:\penalty0 239--261, May 2004.
\newblock \doi{10.1016/j.aquaculture.2003.11.029}.
\newblock URL
  \url{https://www.sciencedirect.com/science/article/pii/S0044848603008007}.

\bibitem[Pecherska(2019)]{pecherska_us_2019}
Tetyana Pecherska.
\newblock U.{S}. {Offshore} {Aquaculture} {Potential} in the {Atlantic} {Ocean}
  and {Gulf} of {Mexico}.
\newblock In \emph{{AAG} {Annual} {Meeting} 2019}, April 2019.
\newblock URL
  \url{https://aag.secure-abstracts.com/AAG%20Annual%20Meeting%202019/abstracts-gallery/22695}.

\bibitem[{Statista}(2024)]{statista_salmon_2024}
{Statista}.
\newblock Salmon price index worldwide 2023, March 2024.
\newblock URL
  \url{https://www.statista.com/statistics/1195271/price-salmon-price-index/}.

\end{thebibliography}

\end{document}